\documentclass[twocolumn,prb,aps,amsfonts]{revtex4-1}
\usepackage{latexsym}
\usepackage{dcolumn}
\usepackage[dvips]{graphicx}
\usepackage{amssymb}
\usepackage{graphicx}
\usepackage{psfrag, color}
\usepackage{verbatim}
\usepackage{amsmath}
\usepackage{epsf}

\begin{document}


\title{The apparent 2D metal-insulator transition as a strong localization
  induced crossover phenomenon: Implications from the Ioffe-Regel criterion}  
\author{S. Das Sarma$^1$ and E. H. Hwang$^{1,2}$}
\address{$^1$Condensed Matter Theory Center, 
Department of Physics, University of Maryland, College Park,
Maryland  20742-4111 \\
$^2$SKKU Advanced Institute of Nanotechnology and Department of
Physics, Sungkyunkwan 
University, Suwon, 440-746, Korea} 
\date{\today}

\begin{abstract}
Low-disorder and high-mobility two-dimensional (2D) electron (or hole)
systems confined in semiconductor heterostructures undergo an apparent
metal-insulator-transition (MIT) at low temperatures as the carrier
density ($n$) is varied. In some situations, the 2D MIT can be caused
at a fixed low carrier density by changing an externally applied
in-plane magnetic field parallel to the 2D layer. 
The goal of the current work is to obtain the critical density ($n_c$)
for the 2D MIT with the system being an effective metal (Anderson
insulator) for density $n$ above (below) $n_c$.
We study the 2D MIT phenomenon theoretically
as a possible strong localization induced crossover process
controlled by the Ioffe-Regel criterion, $k_F l=1$, where $k_F(n)$ is
the 2D Fermi wave vector and $l(n)$ is the disorder-limited quantum
mean free path on the metallic side. Calculating the quantum mean free
path in the effective metallic phase from a realistic Boltzmann transport theory
including disorder scattering effects, we solve the integral equation
(with $l$ depending on $n$ through multidimensional integrals)
defined by the Ioffe-Regel criterion to obtain the nonuniversal
critical density $n_c$ as a function of the applicable physical
experimental parameters including disorder strength, in-plane magnetic
field, spin and valley degeneracy, background dielectric constant and
carrier effective mass, and temperature. The key physics underlying
the nonuniversal parameter dependence of the critical density is the
density dependence of the screened Coulomb disorder. Our
calculated results for the crossover critical density $n_c$ 
appear to be in qualitative and semi-quantitative agreement
with the available experimental data in different 2D semiconductor
systems lending credence to the possibility that the apparent 2D MIT
signals the onset of the strong localization crossover in disordered
2D systems. 
We also compare the calculated critical density obtained from the
Ioffe-Regel criterion with that obtained from a classical percolation
theory, concluding that experiments support the Ioffe-Regel criterion
for the 2D MIT crossover phenomena.
\end{abstract}

\maketitle

\section{introduction}

Carrier transport in 2D semiconductor structures [e.g., Si inversion
  layers in MOSFETs, 2D electron gas (2DEG) or 2D hole gas (2DHG)
  GaAs-AlGaAs heterojunctions and quantum wells or in Si-SiGe quantum
  wells] is a strong function of carrier density ($n$) and temperature
($T$) \cite{andormp,abrahamsrmp,kravchenko,ssc,spivakrmp,dassarmarmp}. 
The study
of 2D electronic (which include holes also in 2D p-doped structures)
transport at low temperatures divides itself naturally into two
distinct areas: effectively ``metallic" transport at high carrier
density manifesting a weak or moderately positive $d\rho/dT \agt 0$
and effective insulating temperature dependence at low carrier density
manifesting a large negative $d\rho/dT <0$ at low temperatures, where
$\rho(T)$ is the temperature dependent 2D resistivity.  
The current work aims at a theoretical understanding of the
density-tuned crossover behavior between this high-density effective
metallic and the low-density effectively insulating transport behavior
at low-temperatures. Consistent with the widely used terminology, we
refer to this density tuned low-temperature phenomenon (i.e., a change
of sign in $d\rho/dT$) as the 2D metal-insulator transition (``2D
MIT"), but it is more likely to be a crossover behavior rather than an
actual quantum phase transition, and indeed in the current work we
treat the 2D MIT phenomenon manifestly as a crossover behavior (as
already made explicit in the title of this paper) characterized by the
Ioffe-Regel criterion. We emphasize right here in the beginning that
ours is not a critical theory for a quantum phase transition. 
This is also highlighted in the title of our paper through the words
``apparent'' and ``crossover''.
We also emphasize that our current work is not a theory for the
temperature-dependent transport properties of the metallic or the
insulating phase, which have been much studied in the literature, but
is a theory for the transition density itself, which has not been
studied at all in the theoretical literature.

Prior to $1979 - 80$, when the scaling theory of localization
\cite{abrahams1979} came into being, the 2D MIT phenomenon in Si
inversion layers (based on Si-SiO$_2$ MOSFET structures) was
universally considered to be an example of the Anderson localization
transition \cite{andormp,mott1975,adkins1978}, where decreasing 2D
carrier density drives the system from being a 2D metal at high
density to being a 2D insulator at lower density as the Fermi level
moves through a mobility edge. The scaling theory of localization
\cite{abrahams1979} established two to be the lower critical
dimensionality in the non-interacting localization problem, and thus
all disordered 2D electronic states are now (i.e., after 1979) thought
to be strictly localized although the localization length is
exponentially long for weak disorder, making it essentially impossible
to distinguish a true extended metallic state from a weakly localized
insulating state in finite samples (or at finite temperatures) with
the localization length exceeding the sample size (or the
temperature-dependent inelastic phase breaking length). For high
disorder, however, the 2D system becomes exponentially localized
(``strong localization") and the crossover from ``weak localization"
(which is a logarithmically weak effect) to ``strong localization"
(which is exponential localization with the
single-particle wavefunction falling off exponentially with distance
in contrast to weak localization with exponentially long localization
length) is the subject matter of the current work. With no loss of
generality, the weakly localized higher-density phase will be referred
to as the ``metallic" phase in this paper and the lower density
strongly localized phase will be referred to as the ``insulating"
phase. Experimentally, the two phases (i.e., high-density
weakly-localized or ``metallic" and low-density strongly localized or
``insulating") are distinguishable by the temperature dependence of
their respective low-temperature resistivity with the strongly
localized low-density phase manifesting the exponential insulating
behavior with a strongly exponentially increasing resistivity with
decreasing temperature whereas the metallic high-density phase either
exhibiting almost no temperature dependence in its resistivity (at
very high density) or the resistivity decreasing (as an effective
power law) with decreasing temperature (at not very high density). In
principle, the "metallic" phase should manifest a weak logarithmic
insulating temperature   
dependence of low enough temperatures because of the weak localization
property of the 2D metallic phase as predicted by the scaling theory,
but the observation of such a weak logarithmic insulating temperature
dependence is challenging since it is easily overwhelmed by any other
power-law temperature dependence in the system (except perhaps at
extremely low temperatures) arising, for example, from phonon or
screening effects. At very low temperatures, the logarithmic
insulating temperature dependence should dominate even in high-density
samples, but such low electronic temperatures are often impossible to
reach because of carrier heating problems invariably present in
semiconductors. We ignore the 2D
weak localization complications \cite{weak1,weak2} in the rest of this
paper, referring 
to the higher-density phase as an effective 2D metal (and the
lower-density strongly localized phase as a 2D insulator).

In this paper we theoretically study the weak to strong localization
transition (or the apparent transition from the effective metal to the
effective insulator) as a function of carrier density at a fixed low
temperature. It is expected that this transition, which is technically a
crossover, happens around $k_F l = 1$ as defined by the so-called
Ioffe-Regel criterion, where $k_F$ is the 2D Fermi wave vector and
`$l$' is the disorder-induced elastic quantum mean free path. We note
that $k_F \propto \sqrt{n}$ in 2D systems, but `$l$' itself is a
complicated functional of `$n$' defined through a complex integral
function $l(n)$. Thus, the condition $k_F l =1$ would define an
effective crossover ``critical density" $n_c$ with $n>n_c$ being the
2D metallic phase (where $l>k_F^{-1}$) and $n<n_c$ being the (strongly
localized) 2D insulating phase (where $l <k_F^{-1}$ in our theory). We
emphasize that the $n>n_c$ metallic phase is only an effective metal
in finite samples.

In addition to obtaining the qualitative dependence of the crossover
density $n_c$ for 2D MIT as implied by the Ioffe-Regel criterion on
the 2D materials parameters (e.g., effective mass, valley degeneracy,
dielectric constant,, etc.), our main goal would be to ascertain the
qualitative dependence of $n_c$ on disorder. Since disorder determines
the mean free path `$l$', the crossover `critical' density $n_c$, as
obtained from the condition $k_Fl=1$, would depend crucially on
disorder parameters. More disordered the system, the higher would be
$n_c$ since the effective mean free path would be smaller for larger
disorder. Since the effective disorder (and therefore the mean free
path itself) is temperature dependent, $n_c$ would manifest an
implicit temperature dependence through the temperature dependence of
the 2D metallic conductivity, which we would also study
theoretically although, strictly speaking, $n_c$ is defined only at
$T=0$ which is indeed the focus of our theory.
Another aspect of experimental interest we study
theoretically in this paper is the effect of a parallel magnetic field
which tends to spin-polarize the system leading to suppressed
screening and hence enhanced effective disorder (and thus a suppressed
metallic conductivity and a reduced mean free path). Thus, an in-plane
applied magnetic field increases $n_c$ (since it decreases the
effective mean free path) consistent with experimental observations.  
The dependence of the crossover critical density $n_c$ on
disorder and magnetic field is the focus of our
current theoretical work, where we uncritically use the Ioffe-Regel
criterion to define the metal-to-insulator apparent transition.
We emphasize that this work is not a theory describing  the properties
of either the higher-density ($n>n_c$) apparent metallic or the
lower-density ($n<n_c$) insulating phase. These phases have already been
studied extensively in the literature over the last 20 years
\cite{andormp,abrahamsrmp,kravchenko,ssc,spivakrmp,dassarmarmp}, 
and we have little to add to these discussions in the current work
which is focused entirely on describing the properties of the
crossover critical density (i.e., $n_c$) itself using the Ioffe-Regel
criterion as the underlying theoretical principle.

It is important to emphasize that our work is purely phenomenological
in nature where the Ioffe-Regel criterion, $k_F l =1$, plays the
central role in determining the crossover density for 2D MIT. Whether
this specific description for 2D MIT is valid or not can only be
decided {\it a posteriori} by comparing between our calculated
theoretical results for $n_c$ 
and the corresponding experimental results. We
should mention right now that our calculated $n_c$ appears to be in
reasonable qualitative and semi-quantitative agreement with the
available experimental results (as would later be discussed in this
article) on 2D MIT in the literature although detailed quantitative
comparisons are difficult since independent experimental information
on the applicable disorder in the relevant 2D systems is
unavailable. We assume in this work that the dominant disorder in the
semiconductor structures undergoing 2D MIT arises from uncorrelated
random quenched charged impurities in the environment
\cite{klapwijkssc1999}. There is considerable experimental evidence
\cite{andormp,hwangprb2008,lewalleprb2002,lillyprl2003, senzpe2002,
  tracyprb2009, manfraprl2007, nohprb2003} that unintentional random
charged impurities in the background and at the interface as well as
the remote charged donors (e.g., in modulation doped heterostructures
and quantum wells) are the main resistive scattering sources in 2D
semiconductor systems at low temperatures (and at low carrier
densities where 2D MIT occurs). In Si-MOSFETs the short-range disorder
associated with Si-SiO$_2$ interface \cite{andormp} certainly plays a
role in resistive scattering at higher carrier density (when the 2D
electron gas is pushed very close to the interface by the
self-consistent electric field generated by the electrons themselves),
but for $n \agt n_c$ the main scattering source is the charged
impurity disorder even for Si-MOSFETs provided that $n_c$ is not too
large \cite{klapwijkssc1999}. We therefore neglect all disorder
mechanisms other than random charged impurities in the system, which
we parameterize using only two parameters: $n_i$ and $d$ [a 2D density
  of charged impurity centers of concentration $n_i$ distributed
  randomly in a plane a distance `$d$' away from the
  semiconductor-insulator interface where the 2D carrier system is
  localized -- we assume the charged impurities to be of random
  positive and negative sign of unit strength (i.e., of magnitude $e$,
  the electron charge) with a net charge of zero].  
We emphasize that it is straightforward to include in the theory
additional (as well as more realistic) types of disorder (e.g.,
interface roughness, alloy disorder, bulk 3D distribution of random
charged impurities) and/or a more general model for charged impurity
disorder where some correlations among impurity positions are included
in the theory \cite{liprl2011}. We neglect all these non-essential
details simply because including them would require many more
additional unknown parameters to characterize the system disorder
compared with our minimal model of disorder which is characterized by
only two parameters $n_i$ and $d$. Since our minimal model
(characterized by only two parameters $n_i$ and $d$) already captures
the essential physics of 2D MIT,
we believe that this minimal model should suffice for our
theoretical purpose. 
In addition to providing results for $n_c$ based on the quantum
Ioffe-Regel criterion, we also carry out a comparison between the
critical densities obtained from the Ioffe-Regel theory and from the
classical percolation theory which has sometimes been invoked in
describing 2D MIT in the context of long-range Coulomb disorder. A
comparison between $n_c$ calculated in these two theories and the
experimental $n_c$ should tell us whether 2D MIT is a quantum or
classical phenomenon.

The rest of this article is organized as follows. In section II 
we provide the details for the calculation of the crossover critical
density $n_c$ using the Ioffe-Regel criterion, giving the general
theory, the analytical results, and all the numerical results for
realistic systems. We provide a general discussion of our results in
section III, particularly in the context of Ioffe-Regel versus
percolation criteria as descriptions for the 2D MIT phenomena. We
conclude in section IV with a summary of our results and the
discussion of open questions.

\section{Calculation of critical density}

In this section, we apply the Ioffe-Regel criterion
\begin{equation}
k_F l=1,
\label{kfl}
\end{equation}
to calculate the crossover 2D MIT `critical' density $n_c$ as a function of
system parameters. We first note an immediate problem 
arising from the uncritical direct application of Eq.~(\ref{kfl}) in
obtaining the critical density $n_c$. Interpreting `$l$' as the
transport (or conductivity) mean free path we can write 
\begin{equation}
l = v_F \tau_t,
\label{mfp}
\end{equation}
where $v_F$ and $\tau_t$ are respectively the Fermi velocity and the
transport relaxation time. Using the well-known Drude-Boltzmann
formula for the electrical conductivity $\sigma$ given by 
\begin{equation}
\sigma = ne^2 \tau_t/m,
\label{sigma}
\end{equation}
and $k_F = (4\pi n/g_s g_v)^{1/2}$ for the 2D electron gas (with $g_s$
and $g_v$ being respectively the ground state spin- and
valley-degeneracy factor) and the
definition $v_F = p_F/m = \hbar k_F/m$, we get from Eq.~(\ref{kfl})
the following condition for the critical conductivity $\sigma_c =
\sigma(n_c)$  at the transition
\begin{equation}
\sigma_c =  \frac{g_s g_v}{2}  \frac{e^2}{h}.
\label{sigmac}
\end{equation}
Equation ({\ref{sigmac}), which is precisely equivalent to the
  Ioffe-Regel criterion of Eq.~({\ref{kfl}) for a 2D electron system,
    can also be written as  
\begin{equation}
\rho_c \equiv \sigma_c^{-1}= \frac{2}{g_sg_v} \frac{h}{e^2}.
\label{eq18}
\end{equation}
Thus, in 2D systems the Ioffe-Regel criterion is precisely equivalent
to $\rho_c = h/e^2$ if we take the usual situation of $g_s=2$ and
$g_v=1$ whereas, for Si(100)-MOSFETs with $g_v=2$, we get $\rho_c =
h/2e^2$. It is interesting to note that the straightforward
application of Ioffe-Regel criterion leads to a critical metallic
resistivity of only $h/6e^2 \sim 4,400 \; \Omega$ for the 6-fold
valley degenerate Si(111)-MOSFETs which have recently been fabricated
using Si-vacuum interfaces \cite{kane111}.

Such a universal critical resistivity characterizing 2D MIT, with
$\rho_c=h/e^2 \approx 25,600 \; \Omega$ for 2D n-GaAs and p-GaAs and
$\rho_c \approx 12,800 \; \Omega$ for Si(100)-MOSFETs [or $4,400 \;
  \Omega$ for Si(111)-MOSFETs], is, however, in quantitative
disagreement with experimental observations where the reported critical
resistivity for the insulating behavior to manifest itself is
certainly not universal in a single material system (i.e., for a given
$g_s$ and $g_v$) and typically varies between $10,000 \; \Omega$ and
$50,000 \; \Omega$ in various experimental studies although it is
often typically within a factor of two of the resistance quantum
$h/e^2$, 
thus indicating that the naive consideration given by the direct
application of the Ioffe-Regel criterion defined by
Eqs.~(\ref{sigmac}) and (\ref{eq18}) is certainly reasonably, but not
perfectly, accurate. 
In fact, the pioneering experimental studies of Kravchenko {\it et
  al}. \cite{kravchenko1995} found $\rho_c \approx 1.5 \; h/e^2$ in
low-disorder MOSFETs and the older highly disordered MOSFETs
\cite{andormp} typically manifested $\rho_c \approx h/4e^2$ although
both classes of systems presumably involved $g_v=2$ and $g_s=2$ with
the only difference between the two being the level of disorder and
the concomitant value of $n_c$ (being around $\sim 10^{11}$ cm$^{-2}$
and $\sim 10^{12}$ cm$^{-2}$ in two classes of systems,
respectively). 
The measurement temperature used in the experiment is also a
complication, particularly since the measured resistivity is typically
strongly temperature dependent for $n \approx n_c$.
The most recent experimental investigations of 2D MIT
in high quality Si(100) MOSFET based and Si-Ge based 2D systems (both
should have $g_s=g_v = 2$) manifest $\rho_c \approx 2 h/e^2$ and
$2h/3e^2$, respectively, in contrast to the canonical value $h/2e^2$
[Eq.~(\ref{eq18})] expected on the basis of $g_s=2$ and $g_v=2$. In
n-2D GaAs systems, $\rho_c \approx h/2e^2$ has been found
\cite{lillyprl2003} whereas in 2D p-GaAs, the observed $\rho_c$ seems
to vary widely with $\rho_c \approx 2h/e^2$ \; \cite{manfraprl2007}
and $\rho_c \approx h/2e^2$ \; \cite{nohprb2003}  both being reported
in contrast to the theoretically expected $\rho_c = h/e^2$ for
$g_s=2$, $g_v=1$ 2D systems. 

Thus, we have a conundrum in using the Ioffe-Regel criterion with
`$l$' interpreted as the transport mean free path since this would
lead to a (upto a factor of 2) quantitative inconsistency with the
experimentally observed 
variations in $\rho_c \equiv \rho(n_c)$ at the 2D MIT crossover
point. We note that this problem of a non-universal experimental
$\rho_c$ for 2D MIT within the same material system (i.e., constant
$g_s$ and $g_v$) in contrast with the universal theoretical $\rho_c$
prediction from the Ioffe-Regel criterion (with `$l$' interpreted as
the transport mean free path) cannot be resolved by altering the
criterion to a different form such as the Mott-Ioffe-Regel criterion
\cite{grahamjpc} where the transition is defined by $k_F l = \pi$
     [rather than $k_Fl=1$ as in the original Ioffe-Regel condition
       defined by Eq.~(\ref{kfl})]. Such a modification will only
     alter Eq.~(\ref{eq18}) for $\rho_c$ to $\rho_c = (2/\pi g_sg_v)
     (h/e^2)$, a different (and smaller) universal critical
     resistivity for constant values of $g_s$ and $g_v$ which is, of
     course, still in disagreement with the experimental
     observations. 
Our possibility that cannot be ruled out in this context is that the
critical resistivity $\rho_c$ at $T=0$ indeed satisfies the expected
resistance quantum value given by Eq.~(\ref{eq18}), but cannot really be
accurately determined from finite temperature crossover
measurements. The fact that the experimentally measured $\rho_c$ is
almost always within a factor of 2 of the expected value defined by
Eq.~(\ref{eq18}) makes this possibility particularly relevant.

A simple modification of the Ioffe-Regel criterion, where `$l$' is
interpreted as the quantum mean free path given by $l=v_F \tau_q$
where $\tau_q$ is the quantum single-particle scattering time (rather
than the transport relaxation time $\tau_t$), actually provides a
variable critical resistivity since there is no simple relationship
connecting the conductivity $\sigma$ with the quantum scattering time
$\tau_q$. Using the identity that impurity scattering induced quantum
level broadening $\Gamma$ is related to $\tau_q$ by  
\begin{equation}
\Gamma = \hbar/2 \tau_q,
\end{equation}
it is easy to see that the Ioffe-Regel criterion Eq.~(\ref{kfl}),
based on using $l = v_F \tau_q$, becomes 
\begin{equation}
\Gamma = E_F,
\label{eq20}
\end{equation}
where $E_F = \hbar^2 k_F^2/2m = (\hbar^2/2m)(4\pi n g_s g_v)$.
We will use this modified Ioffe-Regel criterion in some of our
theoretical analyses since this condition implies a non-universal
critical resistivity $\rho_c$ at the 2D MIT crossover even for the
same values of $g_s$ and $g_v$. 
We emphasize, however, that most of our results are derived based on
the standard Ioffe-Regel criterion where $l$ is taken as the transport
mean free path.
We mention that the condition defined by Eq.~(\ref{eq20}), which arises
from assuming $k_F v_F \tau_q = 1$, is meaningful since extended
metallic states described by momentum eigenstates cannot exist when
the level broadening becomes equal to the Fermi energy since the
momentum is then no longer a good quantum number. In ordinary metals,
one always has $\tau_q=\tau_t$ and hence this issue becomes irrelevant.

We note, however, that the calculation of the crossover critical
density $n_c$ itself, either using the transport mean free path or the
quantum mean free path, would give similar qualitative (but different
quantitative) results. Since the precise sample disorder is never
quantitatively known (and since we use approximations in treating
disorder scattering effects), our goal in this work is a qualitative
(and not quantitative) evaluation of the dependence of $n_c$ on
various physical variables such as disorder (i.e., $n_i$ and $d$),
mobility (at high density), temperature, magnetic field, and materials
parameters (e.g., $g_v$, $g_s$).

We note that in most systems where the Ioffe-Regel criterion has been
applied and discussed so far in the literature (see Graham {\it et
  al.} \cite{grahamjpc} and references therein) there is virtually no
difference between the transport relaxation time $\tau_t$ and the
quantum scattering time $\tau_q$ since the effective disorder
potential is short-ranged. In high-mobility
modulation-doped 2D systems, however, the charged dopants are placed
far from the plane of the 2D layer where the carriers (either
electrons or holes in n- or p-modulation doped GaAs or Si-Ge quantum
wells and heterostructures) are located, leading to an essentially
unscreened very long-range disorder potential in the 2D system. In
such modulation-doped high-mobility 2D systems, it is possible for
$\tau_t \gg \tau_q$ since most of the disorder scattering would be
forward scattering, suppressing $\tau_q$ without affecting
$\tau_t$. \cite{dassarmaprb1985} In such a situation, where forward
scattering by remote dopants ($k_Fd \gg 1$) dominates transport, it is
possible for $\rho_c$ given by Eq.~(\ref{eq20}) to be smaller than the
$\rho_c$ defined by Eq.~(\ref{eq18}). Consequently, the crossover
critical density $n_c$ will then be higher as given by
Eq.~(\ref{eq20}) with $\Gamma = \hbar/2\tau_q$ compared with that
given by Eq.~(\ref{kfl}) with $l=v_F \tau_t$ defined by the transport
mean free path. For Si-MOSFETs, most of the disorder is of
short-ranged nature (either screened charged impurities near the interface or
surface roughness scattering), and therefore, $\tau_t \approx \tau_q$,
so that Eqs.~(\ref{kfl}) and (\ref{eq20}) should give similar (but not
identical) estimates for $n_c$ and $\rho_c$ with 
\begin{eqnarray}
n_c \;\; [{\rm obtained \; by \; Eq.~(\ref{eq20})}] & \geq & n_c \;\;
[{\rm obtained \; by \; Eq.~(\ref{kfl})}]  \nonumber \\ 
\rho_c \;\; [{\rm from \; Eq.~(\ref{eq20})}] & \leq & \rho_c \;\;
    [{\rm from \; Eq.~(\ref{eq18})}]. 
\label{eq21}
\end{eqnarray}
The inequalities given in Eq.~(\ref{eq21}) above are general applying
to all 2D and 3D systems, and follow simply from the fact that $\tau_q
\alt \tau_t$ always.

\subsection{Theory}

\subsubsection{Disorder dependence}
\label{sub31}

Starting with Eq.~({\ref{kfl}) and writing $l = v_F \tau$ where
  $\tau$ is an impurity-induced scattering time (either $\tau_t$ or
  $\tau_q$), we can derive the following scaling relation 
\begin{equation}
n_c \sim n_i^{\gamma}
\label{eq22}
\end{equation}
where
\begin{equation}
\gamma=(1+\delta)^{-1}
\end{equation}
and the exponent `$\delta$' defines the density dependence of $\tau$
\begin{equation}
\tau \sim n^{\delta}.
\label{eq24}
\end{equation}
In deriving Eq.~(\ref{eq22}), we assume that all parameters, other
than $n_i$, are fixed and disorder is entirely parameterized by the 2D
impurity density $n_i$. 
There is an implicit dependence of $n_c$ on the background dielectric
constant $\kappa$ and on the impurity location parameter $d$ not
explicitly shown in  Eqs.~(\ref{eq22}) -- (\ref{eq24}). We assume (at
this stage) that the $d$-parameter (and the dielectric constant
$\kappa$) characterizing the samples is approximately a constant so
that the disorder strength can be  characterized by the single
parameter 2D impurity density $n_i$ in Eq.~(\ref{eq22}). 

The sample mobility itself is, by definition, inversely proportion to $n_i$
\begin{equation}
\mu \sim n_i^{-1},
\end{equation}
enabling us to eliminate $n_i$ in Eq.~(\ref{eq22}) in favor of some
``maximum mobility" $\mu_m$ defined at a high fiduciary carrier
density $n_m \gg n_c$. Eliminating $n_i$ in favor of $\mu_m^{-1}$ we
get 
\begin{equation}
n_c \sim \mu_m^{-\gamma}.
\label{eq26}
\end{equation}
This gives us a scaling relationship connecting the crossover density
$n_c$ to the sample quality as characterized by the typical ''maximum
mobility" $\mu_m$ defined at some high carrier density $n_m \gg n_c$
deep in the metallic phase.

A detailed theory has recently been developed by us
\cite{dassarmaprb2013} for the density scaling of 2D metallic
conductivity (and mobility), where we find that the exponent $\delta$
(with $\mu \sim n^{\delta}$) given by 
\begin{equation}
\delta \approx 0.7 \; ({\rm n-GaAs}); \; 0.5 \; ({\rm p-GaAs}); \; 0.3
\; ({\rm n-Si}), 
\label{eq27}
\end{equation}
restricting to the $n \gg n_c$ situation. This then implies
\begin{equation}
\gamma = 1/1.7 - 1/1.3 \approx 0.59 - 0.77,
\end{equation}
with 
\begin{equation}
\gamma \approx 0.59 \;({\rm n-GaAs}), \; 0.67 ({\rm p-GaAs}); \; 0.77
({\rm n-Si}). 
\label{eq29}
\end{equation}
We emphasize that Eqs.~(\ref{eq27}) -- (\ref{eq29}) are very
approximate and are derived assuming that the 2D impurity density
$n_i$ is the only variable determining the disorder, and therefore the
quantitative applicability of the numerical values of the exponent
$\gamma$ (defining $n_c \sim \mu_m^{-\gamma}$) is very approximate. 
We note that for purely short-range $\delta$-function scatterers, we
get $\delta=0$ and $\gamma=1$, i.e., $n_c \sim n_i$ for purely
zero-range disorder.

It is, therefore, important to emphasize that such a scaling
relationship, $n_c \sim \mu_m^{-\gamma}$, with $\gamma \approx 0.67$
approximately (but with some fluctuations in the distribution of
$\gamma$ values around $\gamma \sim 0.67$) was noted empirically by
Sarachik \cite{sarachikepl} more than ten years ago based on a careful
numerical analysis of the existing 2D MIT experimental data. Thus, our
Ioffe-Regel criterion based theoretical analysis of the dependence of
the crossover critical density $n_c$ on the typical sample mobility
$\mu_m$ is consistent with the 2D MIT experimental data
\cite{sarachikepl}. 
This agreement between the experimental mobility dependence of $n_c$
and our Ioffe-Regel criterion based results is one of the main
{\it a posteriori} justifications for our theory.

Similar theoretical considerations can also be applied to the case
where $\tau$ is interpreted to be the quantum scattering time $\tau_q$
[rather than the transport scattering time $\tau_t$ as in the analysis
  of Eqs.~(\ref{eq22}) -- (\ref{eq29}) above]. For this situation the
Ioffe-Regel criterion is better written as $\Gamma = E_F$ [see
  Eq.~(\ref{eq20})], and it is straightforward to show that we get the
following results for $n_c$ (where $q_{TF}$ is the 2D Thomas-Fermi
screening wave vector \cite{andormp})
\begin{equation}
n_c = n_i q_{TF}^2 \int_0^1 \frac{dx}{\sqrt{1-x^2}}
\frac{e^{-4k_Fdx}}{(x+q_{TF})^2}, 
\label{eq30}
\end{equation}
which leads immediately to
\begin{equation}
n_c =\frac{1}{4 \sqrt{2\pi}} \left ( \frac{n_i}{d} \right )^{2/3} \;\;
{\rm for} \; k_Fd \gg1, 
\end{equation}
and
\begin{equation}
n_c = \frac{\pi n_i}{2} \;\; {\rm for} \; k_Fd \ll 1.
\label{eq32}
\end{equation}
Assuming the impurity separation parameter `$d$' to be fixed and the
impurity density $n_i$ to be the sole determinant of the system
mobility, we can adapt Eqs.~(\ref{eq30}) -- (\ref{eq32}) to provide a
dependence of the critical density $n_c$ on some fiduciary maximum
mobility $\mu_m$ (defined as the sample mobility at some high
characteristic density) 
\begin{equation}
n_c \sim \mu_m^{-0.67} \;\; {\rm for} \; k_F d \gg 1,
\end{equation}
and
\begin{equation}
n_c \sim \mu_m^{-1} \;\; {\rm for} \; k_F d \ll 1.
\end{equation}
This immediately leads to the same conclusion we already discussed
above [see the discussion above following Eq.~(\ref{eq29})]  that the
dependence of $n_c$ on the sample quality follows an approximate power
law $n_c \sim \mu_m^{-\gamma}$ where $\gamma \approx 0.5-0.8$, 
as has been already pointed by Sarachik based on an empirical analyses
of the experimental data\cite{sarachikepl}. 
Thus, both the original Ioffe-Regel criterion $k_Fl = 1$ and our
modified Ioffe-Regel criterion $E_F=\Gamma$ give the same theoretical
dependence of $n_c$ on the sample mobility, which is in agreement with
the existing experimental data on 2D MIT.
We note that a pure short-range disorder model on the other hand gives
the exponent $\gamma=1$ disagreeing with the empirical data on 2D MIT.

\subsubsection{Temperature dependence}
\label{sub32}

The Ioffe-Regel criterion strictly applies at $T=0$, but can be
generalized to finite temperatures by considering the temperature
dependence of the mean free path `$l$' on the metallic side. This is,
of course, relevant for the 2D MIT problem since its
hallmark (and the {\it raison d'etre} for its huge impact in contrast
to the corresponding MIT phenomenon in 2D semiconductor systems in the
$1970-90$ era) is the strong temperature dependence of the 2D metallic
conductivity for $n \agt n_c$. The strong temperature dependence of
the 2D metallic conductivity for $n \agt n_c$ immediately leads to a
strong temperature dependence of the mean free path $l(T)$, which
should affect the critical density $n_c(T)$ derived on the
basis of the Ioffe-Regel criterion $k_Fl=1$. Since the 2D metallic
conductivity decreases with increasing temperature for $n \agt n_c$,
the corresponding $l(T)$ also decreases with increasing temperature
whereas the Fermi wave vector $k_F \propto \sqrt{n}$ is, by
definition, temperature-independent.  
This implies that $n_c(T)$, defined by the Ioffe-Regel criterion,
increases with increasing temperature at the lowest temperatures,
where strong metallicity is observed in high-quality 2D systems. 
The situation is, however, complicated by the fact
\cite{sarmaprb2004,sarmaprb2003} 
that the 2D conductivity is
non-monotonic as a function of temperature, and eventually decreases
with increasing temperature for $T \agt T_F$ which is reached at
pretty low temperatures if $T_c = T_F(n_c)$ is low. Thus, $n_c(T)$
could manifest non-monotonic behavior as a function of temperature for
a given sample with $n_c(T)$ increasing with $T$ at the lowest
temperatures, and then decreasing with $T$ at higher
temperatures. Using the
Ioffe-Regel criterion and the Boltzmann transport theory at finite
temperatures \cite{sarmaprb2004,sarmaprb2003}, we get for $T \ll T_c$ 
\begin{equation}
n_c(T) \approx n_c\left [ 1 + \left (\frac{x_0}{1+x_0} \right )
  \frac{T}{T_c} \right ], 
\label{eq35}
\end{equation}
and for $T \agt T_c$
\begin{equation}
n_c(T) \sim \frac{T_c}{T},
\label{eq36}
\end{equation}
where $n_c = n_c(T=0)$; $T_c = T_F(n=n_c)$, and $x_0 = q_{TF}/2k_{Fc}$
where $k_{Fc} = k_F(n=n_c)$.  
(We note that the screening wave vector $q_{TF}$ is a constant
independent of carrier density in 2D because of the constant 2D
density of states.) 
From Eqs.~(\ref{eq35}) and (\ref{eq36}), we conclude that $n_c(T)$
would in principle manifest non-universal behavior, but at
sufficiently high temperatures, $n_c(T)$ will decrease with increasing
temperature approximately linearly, {\it i.e.,} $n_c(T>T_c) \sim
1/T$. This $1/T$ decrease in $n_c(T)$ has been experimentally
observed\cite{lillyprl2005,manfraprl2007,tracyprb2009}, however, the
predicted increase of $n_c(T)$ with $T$ at lower temperatures has not
yet been reported in Si MOSFETs (although it has been seen in 2D p-GaAs holes \cite{manfraprl2007}, 
perhaps because the lowest temperatures are not reached yet in the
experiments to see this decrease or because of complications arising
from weak localization corrections not included in our theory, which
would dominate the asymptotic low-temperature transport.

\subsubsection{Magnetic field dependence}
\label{sub33}

One of the most important experimental discoveries \cite{magnetic} in
the 2D MIT phenomena is the observation of a strong enhancement of
$n_c$ by the application of an applied parallel (to the 2D layer)
magnetic field, which is equivalent to the suppression of the 2D
effective metallic phase by the applied parallel field. In addition to
the enhancement of $n_c$ compared with its zero-field value, an
applied parallel magnetic field also leads to a suppression of the
metallic temperature dependence which becomes weaker as the applied
field is made stronger. While this latter effect of the magnetic field
induced weakening of the metallic temperature dependence has been
extensively studied theoretically\cite{hwangmagnetic}, there has been
no theoretical analysis in the literature of the magnetic field
induced enhancement of $n_c$ itself.

The Ioffe-Regel criterion provides a natural explanation for the field
induced enhancement of $n_c$ as arising from the suppression of the
metallic mean free path `$l$' (or the enhancement of the quantum level
broadening $\Gamma$) due to the enhancement of the effective
disorder in the metallic phase as the effective carrier screening is
reduced by the application of the applied parallel magnetic field
$B$. Screening is suppressed at finite $B$ since the system gets spin
polarized by the $B$-induced Zeeman splitting so that the effective
Thomas-Fermi screening wave vector $q_{TF}$ defined by 
\begin{equation}
q_{TF} = \frac{g_sg_v me^2}{\kappa \hbar^2},
\label{eq37}
\end{equation}
decreases as the spin degeneracy changes from $g_s=2$ for $B=0$ to
$g_s=1$ for $B=B_s$ where $B_s$ is the density dependent field
strength that completely spin-polarizes the 2D system, i.e., $B_s$ is
defined by 
\begin{equation}
2 g \mu_B B_s = E_F,
\end{equation}
where $g$ is the Land\'{e} $g$-factor for the specific semiconductor,
$\mu_B$ is the Bohr magneton, and $E_F$ is the Fermi energy at
$B=0$. For $B=B_s$, the 2D system is completely spin-polarized at the
Fermi level with $g_s=1$. We note, however, that $k_F=(4\pi
n/g_sg_v)^{1/2}$ itself is now $B$-dependent in the presence of
spin-polarization, and becomes $\sqrt{2}$ times larger at $B=B_s$
compared with its value at $B=0$ since $g_s$ decreases from 2 to
1. Thus, if the mean free path $l$ remains unaffected by the applied
field $B$, then the effect of spin-polarization-induced enhancement of
$k_F$ itself would lead to a decrease of $n_c$ in finite $B$ in
apparent disagreement with experimental observations. Thus,
spin-polarization-induced (or equivalently $B$-induced) suppression of
screening (and the consequent enhancement of Coulomb disorder) is
essential in understanding the enhancement of $n_c$ in the presence of
finite applied parallel field. If the effective disorder underlying
the 2D MIT phenomenon is purely short-ranged $\delta$-function
white-noise disorder, where carrier screening should play no key role,
then the application of the parallel magnetic field would decrease
$n_c$, effectively enhancing the metallic phase rather than
suppressing it as observed experimentally. This latter effect (but not
the former), i.e., the effect of the enhancement of $k_F$ by
spin-polarization (but not the effect of suppressed screening), is
already implicit in Eq.~(\ref{eq22}) which implies a decreasing $n_c$
with  decreasing $g_s$ (i.e., with increasing applied field). 
We mention that for very large $n_c$, so that $q_{TF}/2k_F$ is very
small and screening is unimportant, we do predict that an applied
parallel field will either have almost no effect on $n_c$ because
spin-polarization effects are negligible due to the very small
spin-polarization induced by an applied field at very large density
(or will decrease it because of the increasing $k_F$ with increasing
spin-polarization).  One reason that the applied field effect on 2D
MIT was not discovered until 2000 is indeed the fact that $n_c$ was
simply too large in the older (and dirtier) 2D samples for the
field-induced screening suppression to play any role.

To include the effect of suppressed screening in the presence of
finite spin-polarization (i.e., $g_s < 2$), we must take into account
the variation of $q_{TF}$ with $g_s$ as shown in Eq.~(\ref{eq37}). We
can obtain an analytical formula by noting that the screened Coulomb
disorder $u(q)$ behaves in the following manner 
\begin{equation}
u(q) = \frac{v(q)}{\epsilon(q)} = \frac{2\pi e^2}{\kappa (q + q_{TF})},
\end{equation}
where $v(q) = 2\pi e^2/\kappa q$ is the unscreened 2D Coulomb
interaction and $\epsilon(q) = 1 + q_{TF}/q$ is the 2D carrier
dielectric screening function. In the strong screening limit ($q_{TF}
\gg k_F$), we can write $u(q) \sim q_{TF}^{-1} \sim g_s^{-1}$, and
this limit enables an analytical calculation by noting that $l^{-1}
\propto n_i u^2 \propto n_i/g_s^2$, which allows us to replace $n_i$
in Eq.~(\ref{eq22}) by $n_i/g_s^2$, producing the following equation
for $n_c(B)$ taking into account dual effects of the enhancement
(suppression) of both $k_F$ ($q_{TF}$) by the applied field compared
with their $B=0$ values 
\begin{equation}
n_c(B) \sim \left ( \frac{n_i}{g_s} \right )^{\gamma},
\label{eq40}
\end{equation}
which is valid in the strong screening ($q_{TF} \gg k_F$) limit. In
the weak screening limit, the spin-polarization dependence in the
screening may be ignored and we get 
\begin{equation}
n_c(B) \sim (g_s n_i)^{\gamma}.
\label{eq41}
\end{equation}
To obtain the explicit $B$-dependence of $n_c(B)$ we need to express
the spin-degeneracy factor $g_s$ as a function of the magnetic field
$B$, which then leads to 
\begin{equation}
n_c(B) \sim \left ( 1 + \frac{B}{2B_s} \right )^{\gamma},
\label{eq42}
\end{equation}
and
\begin{equation}
n_c(B) \sim \left ( 1 - \frac{B}{2B_s} \right )^{\gamma},
\label{eq43}
\end{equation}
respectively for Eqs.~(\ref{eq40}) and (\ref{eq41}). Since the 2D MIT
phenomenon mostly occurs in the strong-screening ($q_{TF} \gg k_F$)
regime, Eq.~(\ref{eq42}) applies to most situations indicating an
increase in the critical density with increasing applied field. 
At high carrier densities where $2k_F \gg q_{TF}$  or in a situation
where short-range scattering dominates transport so that screening is
not relevant, Eq.~(\ref{eq43}) should apply, and thus for highly
disordered samples with large $n_c$ (so that $q_{TF} \ll 2k_F$), it is
conceivable that an applied field may slightly decrease $n_c$.

Before concluding this section we mention that all our considerations
above for the effect of spin polarization $g_s$ (as modulated by the
parallel field $B$) apply equally well to the valley degeneracy
$g_v$ since the combination $g_sg_v$ appears in all physical
quantities. In particular, if $g_v$ could be modified somehow by an
external valley-symmetry-breaking field $A$ (for example, an applied
strain), then Eqs.~(\ref{eq42}) and (\ref{eq43}) will apply to
describe the valley degeneracy dependence of the critical density with
the $A$-field replacing the $B$-field. In both cases, the lifting of
the spin (or valley) degeneracy by an external field would typically
lead to an increasing critical density with increasing field since
most 2D MIT phenomena happen in the $q_{TF} \agt k_F$ strong screening
regime [as characterized by Eq.~(\ref{eq42})]. Indeed, such a
symmetrical situation of increasing $n_c$ with increasing external
symmetry breaking field for either spin or valley degeneracy has been
experimentally observed by Shayegan and his collaborators in the
multivalley AlAs 2D systems \cite{shayegannp}. Our qualitative
findings in this section are in excellent agreement with these
experimental results \cite{shayegannp} showing an equivalence of
increasing $n_c$ with the decrease of $g_s$ or $g_v$. In particular, a
given sample with a fixed carrier density ($n$) is most likely to be
in the insulating phase [i.e., $n<n_c(g_s,g_v)]$ when the 2D system is
maximally polarized to have the minimum possible values of $g_s$ and
$g_v$ as precisely observed by Shayegan and his collaborators
\cite{shayegannp}.

\subsubsection{Materials dependence}
\label{sub34}

To consider how $n_c$ depends on the materials parameters (e.g., $m$,
$\kappa$, $g_s$, $g_v$) of the 2D system we imagine a situation with
fixed bare disorder (i.e., fixed $n_i$ and $d$) while varying only the
materials parameters to see 
how the Ioffe-Regel criterion $k_F l = 1$ is affected. 

Expressing the mean free path $l=v_F \tau$ in terms of the relaxation
time $\tau$, and then using the Boltzmann equation to obtain $\tau$
assuming scattering from random screened charged impurities we get the
following integral equation for $n_c$ from the Ioffe-Regel condition
(at $T=0$)  $k_F l = 1$ 
\begin{eqnarray}
\frac{1}{\tau} & = & \frac{\hbar k_F^2}{m}  \nonumber \\
                       & = & \frac{n_i m}{\pi \hbar^3 k_F^2} \left (
\frac {2\pi e^2}{\kappa} \right )^2 \int_0^1 \frac{dy}{\sqrt{1-y^2}}
\frac{y^2 e^{-2y d_0}}{(y+x)^2}, 
\label{IRc}
\end{eqnarray}                       
where $x = q_{TF}/2k_F$ and $d_0 = 2 k_F d$.
All other quantities in Eq.~(\ref{IRc}) are defined with $k_F = (4 \pi
n /g_sg_v)^{1/2}$ and $q_{TF} = g_s g_c me^2/\kappa \hbar^2$. We can
rewrite the integral equation defined by Eq.~({\ref{IRc}) as 
\begin{equation}
\pi \hbar^4 k_F^4 = n_i m^2 \left ( \frac{2\pi e^2}{\kappa} \right )^2
\int_0^1 \frac{dy}{\sqrt{1-y^2}} \frac{y^2 e^{-2 d_0 y}}{(y+x)^2}. 
\label{IRc1}
\end{equation}
We note that Eq.~(\ref{IRc1}) is a nontrivial integral equation for
$n_c$ since the density $n$ enters the equation in three distinct
places: $k_F \propto \sqrt{n}$, $d_0 = 2k_F d$, and $x = q_{TF}/2k_F
\propto n^{-1/2}$.  
Before discussing the materials dependence of $n_c$ implied by
Eq.~(\ref{IRc1}), we note that the above relationship [i.e.,
  Eq.~(\ref{IRc1})] has been derived by assuming `$l$' to be the
transport mean free path, i.e., $\tau=\tau_t$. If, instead of $\tau_t$
we use the quantum relaxation time $\tau = \tau_q$, the only
difference is that the vertex correction term disappears from the
integral on the right hand side, leading to the following integral
equation for $n=n_c$ using the $\Gamma = E_F$ Ioffe-Regel criterion
(i.e. $\tau=\tau_q$ in $k_F l = k_F v_F \tau = 1$) 
\begin{equation}
\pi \hbar^4 k_F^4 = n_i m^2 \left ( \frac{2\pi e^2}{\kappa} \right )^2
\int_0^1 \frac{dy}{\sqrt{1-y^2}}\frac{e^{-2d_0y}}{(y+x)^2}, 
\label{IRc2}
\end{equation} 
with the only difference between Eqs.~(\ref{IRc1}) and (\ref{IRc2})
being the additional factor of $y^2$ inside the integral on the right
hand side of Eq.~(\ref{IRc1}) -- this $y^2$ factor arises from the
well-known `$1-\cos \theta$' vertex correction term in the Kubo
formula for the current-current correlation function in the
conductivity.

In Eqs.~({\ref{IRc1}) and (\ref{IRc2}), materials parameters $m$,
  $g_s$, $g_v$, $\kappa$ enter through $k_F = (4\pi n/g_sg_v)^{1/2}$,
  $d_0 = 2k_F d$, $x = q_{TF}/2k_F = (g_s g_v)^{3/2} me^2/(4\kappa
  \hbar^2 \sqrt{\pi n})$, and the factor $2\pi e^2/\kappa$ as well as
  $m$ on the right hand side. It is obvious that no definitive and
  unique dependence of $n_c = n_c(g_v,g_s,m,\kappa)$ on the materials
  parameters can be analytically discerned from Eqs.~(\ref{IRc1}) and
  (\ref{IRc2}) because of the complex functional relationship defined
  by Eqs.~(\ref{IRc1}) and (\ref{IRc2}) of the form $n^2 = n_i
  A(g_v,g_s,m,\kappa) \int_0^1 dy f(y;n,d,g_v,g_s,m,\kappa)$ where
  both $A$ and $f$ are functions of the materials parameters $g_v$,
  $g_s$, $m$, and $\kappa$. The only relatively simple dependence
  implied by the integral equations (\ref{IRc1}) and (\ref{IRc2}) is
  the dependence on the impurity density $n_i$, which has already been
  discussed in subsection \ref{sub31} above in details. We mention
  here that using the Ioffe-Regel criterion the $k_F d \gg 1$ and $k_F
  d \ll1$ limits of Eqs.~(\ref{IRc1}) and (\ref{IRc2}) lead to $n_c
  \sim (n_i/d)^{2/3}$ and $n_c \sim n_i d^0$, respectively.

To discuss the materials dependence of $n_c$ analytically, we start by
assuming that the strong screening ($q_{TF} \gg 2k_F$) condition
applies to the 2D system, which is likely (since $n_c$ is relatively
low for the 2D MIT phenomenon to manifest itself), but not
guaranteed. An additional constraint is necessary on the dimensionless
variable $d_0 = 2k_F d$, which we also assume to be small (which
certainly applies to Si-MOSFETs, but may not always apply to the
modulation-doped GaAs structures even at low values of $n_c$). With
these two constraints (i.e., $q_{TF} \gg 2k_F$ and $2k_F d \ll 1$),
the Ioffe-Regel integral equation can be analytically studied to
provide the following approximate asymptotic dependence of $n_c$ on
materials parameters 
\begin{equation}
n_c \sim (g_s g_v)^{-2/3}; \;\; n_c \sim \kappa^{2/3}; \;\; n_c \sim m^0.
\label{IRc3}
\end{equation}
It is interesting and important to note that in the strong-screening
($q_{TF} \gg 2 k_F$) limit, there is no dependence of the critical
density $n_c$ on the effective mass of the system since the effective
mass appearing in $q_{TF}$ (i.e., screening) exactly cancels out the
inverse effective mass appearing in the Fermi velocity in this
artificial limit $q_{TF} \gg 2k_F$. The spin- and valley-degeneracy
dependence shown in Eq.~(\ref{IRc3}) is the same as what we obtained
in subsection \ref{sub33} above, and indicates that in general
insulating (metallic) phase is preferred by lower (higher) values of
$g_s$ or $g_v$. 
We note, however, that the strong-screening situation itself, $q_{TF}
\gg 2k_F$, depends crucially on the effective mass since $q_{TF} \sim
m$, and thus the strong screening condition is difficult to achieve in
2D systems with small effective mass.

\subsubsection{Comparison with percolation transition}

Ioffe-Regel criterion provides one possibility for
conductor-to-insulator crossover in 2D semiconductor systems. 
Another possibility, which has been studied rather extensively in the 2D
MIT literature \cite{tracyprb2009, manfraprl2007, lillyprl2005,
  percolation,percolation2,efros,barnett}, is a classical
percolation transition at $n=n_c$ arising from the 2D Fermi level
moving through the potential fluctuations (``mountains and lakes"
landscape) associated with the long-range Coulomb disorder in the 2D
system. Simple theoretical considerations \cite{efros} and direct
numerical simulations \cite{barnett} indicate that a 
percolation conductor-to-insulator transition may occur at a critical
carrier density given by 
\begin{equation}
n_c \approx \frac{1}{4 \pi} \frac{\sqrt{n_i}}{d} \approx 0.1
\frac{\sqrt{n_i}}{d}, 
\label{perc}
\end{equation}
where the long-range-Coulomb disorder is created in the 2D layer by
random charged impurities of 2D density $n_i$ located a distance `$d$'
from the 2D layer (i.e., exactly the same model for disorder we have
used in this work for applying the Ioffe-Regel criterion $k_F l =
1$). Since the percolation transition in the context of 2D MIT has
already been extensively studied in the literature, we do not provide
any details in the current work on the 
percolation transition and accept Eq.~(\ref{perc}) for the crossover
density as a given. Our goal in the current work is to compare the
percolation transition with the Ioffe-Regel transition in the context
of the 2D MIT phenomena. It is, however, important here to point out
that the percolation transition is manifestly a classical phenomenon
with the MIT being driven by the chemical potential or the Fermi level
(which is proportional to the carrier density in 2D) crossing through
the percolation point in the potential fluctuation driven
inhomogeneous 2D ``mountains and lakes" landscape with the
high-density ($n \gg n_c$) metallic phase being essentially the
homogeneous (and well-screened) ``all-lakes" conducting situation whereas the
low-density ($n \ll n_c$) insulating phase being the highly
inhomogeneous (and unscreened) ``all-mountains" situation. By
contrast, the Ioffe-Regel  criterion defines a completely quantum
condition for the localization crossover. In some loose sense, the two
criteria (percolation and Ioffe-Regel) are complementary and describe
the 2D MIT as a high-temperature classical and a low-temperature
quantum crossover phenomenon, respectively.

We note that for fixed `$d$', the percolation criterion implies $n_c
\sim \sqrt{\mu_m}$, i.e., the exponent [see. Eqs.~(\ref{eq22}) and
  (\ref{eq26})] $\gamma = 1/2$ in the percolation picture whereas
$\gamma \approx 0.6-0.8$ in the Ioffe-Regel theory as discussed
already in great details above. If we assume $n_i$ to be fixed and
`$d$' to be the relevant variable characterizing impurity disorder,
then the percolation theory gives the simple dependence $n_c \sim
d^{-1}$, which can be converted to the following dependence on the
high-density mobility $\mu_m$ assuming that $k_F d \gg 1$ condition
applies 
\begin{equation}
n_c \sim d^{-1} \sim \mu_m^{-1/3},
\end{equation}
where we have used the fact that 2D mobility $\mu \sim d^3$ for $k_F d
\gg 1$ (and fixed $n_i$). Thus percolation theory gives the following
exponent $\gamma$ (where $n_c \sim \mu_m^{\gamma}$) for the critical
density, assuming $d$ to be fixed,  
\begin{equation}
\gamma = 1/2,
\label{eq51}
\end{equation}
and, assuming $n_i$ to be fixed,
\begin{equation}
\gamma = 1/3.
\label{eq52}
\end{equation}
In addition, the percolation critical density, being dependent only on
$n_i$ and $d$ (i.e., just the bare disorder), is independent of
materials parameters $g_v$, $g_s$, $m$, and $\kappa$ in contrast to
the critical density $n_c$ based on the Ioffe-Regel criterion. We
mention the corresponding Ioffe-Regel exponents for Eqs.~(\ref{eq51})
and (\ref{eq52}) are $\gamma= 2/3$ (fixed $d$) and $2/9$ (fixed $n_i$)
for $k_F d \gg 1$. 
The differences in the exponent $\gamma$ between the two theories are
significant ($\gamma=1/2$ and $1/3$ versus $\gamma = 2/3$ and $2/9$,
respectively), but not very large.

One particular aspect of percolation induced 2D MIT not discussed
above is worth mentioning here (and we will present numerical results
on this aspect later in this paper). The critical density defined by
percolation theory is completely independent of any transport
considerations and thus the constraint on the critical resistivity
$\rho_c \alt (2/g_sg_v)(h/e^2)$ defined by Eqs.~(\ref{eq18}) and
(\ref{eq21}) does not apply to the percolation critical
resistivity. In principle, therefore, $\rho_c$ for the 2D MIT
percolation crossover could be any value much larger or smaller than
the quantum resistance value of $h/e^2$ whereas, by contrast, the
Ioffe-Regel condition implies a critical resistance of $O(h/e^2)$. In
practice, however, we find numerically (as shown in the next section)
that the calculated $\rho_c = \rho(n_c)$ at the 2D MIT percolation
transition turns out to be $\sim h/e^2$ for most, if not all, 2D MIT
experimental parameters in realistic 2D systems. The possibility,
however, remains that $\rho_c$ could be very different from $h/e^2$ in
a percolation 2D MIT crossover since the percolation transition is
simply a classical transition between immobile and mobile states 
depending on the value of the chemical potential (i.e., the Fermi
level) with respect to the disorder potential in an
inhomogeneous potential fluctuation landscape where localization or
quantum interference plays no role. 

While the fact that the percolation critical resistivity $\rho_c =
\rho(n=n_c)$ is, in principle, arbitrary (and can be larger or smaller
than $h/e^2$ with no theoretical constraint) may appear to be an
attractive quantitative feature of the percolation theory in
describing 2D MIT, there are other aspects of the percolation
transition which are in manifest disagreement with the experimental
phenomenology for 2D MIT even on a qualitative level. For example, the
percolation theory predicts an $n_c = n_c(n_i,d)$ which is completely
independent of materials parameters (i.e., $m$, $\kappa$, etc.) and of
the valley and/or spin degeneracy. This is in disagreement with
experimental findings for 2D MIT where, for example, applying an
external in-plane magnetic field to spin-polarize the 2D system leads
to an increasing $n_c$ which cannot easily be described by the
percolation theory.
The classical percolation theory would predict no dependence of $n_c$
on an applied magnetic field.

\subsection{Numerical Results}

We now present detailed numerical results for our calculated critical
crossover density $n_c$ as a function of various physical parameters
using the Ioffe-Regel criterion. These results are obtained by
directly numerically solving the integral equations defined by
Eq.~({\ref{IRc1}) or (\ref{IRc2}), which correspond respectively to
  using $l_t =v_F \tau_t$ or $l_q = v_F \tau_q$ in the Ioffe-Regel
  criterion $k_F l = 1$. Both equations give similar qualitative
  results, and our goal in this work is an investigation of the
  qualitative dependence of $n_c$ on disorder, temperature, applied
  in-plane magnetic field, and system parameters, and we do not
  therefore distinguish between these two closely 
related versions (i.e., $l_t$ or $l_q$) of the Ioffe-Regel
criterion. We also provide a comparison between $n_c^{IR}$ and
$n_c^{per}$ as obtained respectively by the Ioffe-Regel criterion and
percolation transition in same situations. We believe that a direct
comparison between $n_c^{IR}$ and $n_c^{per}$ as a function of
disorder could shed considerable light on the nature of the 2D MIT, in
particular, distinguishing between quantum localization and classical
percolation on a qualitative level.

\subsubsection{Pure 2D case}

In Fig.~\ref{fig1}, we show our numerically calculated critical
density $n_c$ for both the Ioffe-Regel and the percolation theory as a
function of $n_i$ (with $d$ fixed) and $d$ (with $n_i$ fixed). All
numerical results presented in this subsection assume the 2D carriers
to be confined in an ideal strict 2D layer of zero thickness. For the
percolation theory, of course, $n_c = 0.1 \sqrt{n_i}/d$ is trivial to
plot, and we provide these results only for the sake of comparison
with the nontrivial Ioffe-Regel results for $n_c$, which we obtain by
numerically solving the integral equation defined by Eq.~(\ref{IRc2}),
which uses $\tau=\tau_q$ (i.e., $\Gamma = E_F$ Ioffe-Regel
condition). We show results for the three most commonly studied 2D
systems: n-GaAs, p-GaAs, and n-Si(100)-MOSFET (using the appropriate
corresponding values of $m$, $\kappa$, $g_v$, etc. in solving the
integral equation for $n_c$). 

\begin{figure}[t]
	\centering
	\includegraphics[width=1.\columnwidth]{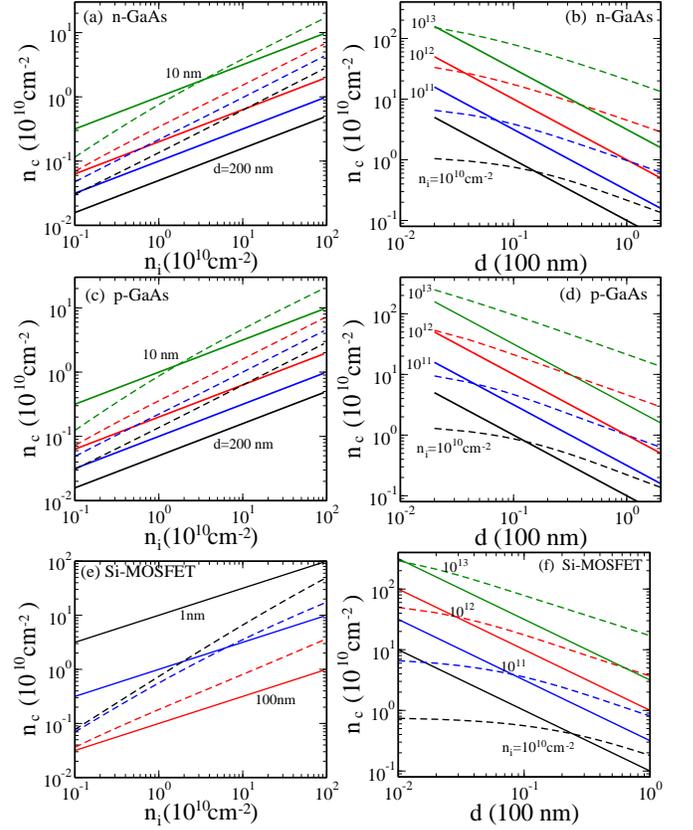}
	\caption{The calculated critical density of n-GaAs (a) as a
          function of  
impurity density $n_i$ for fixed impurity locations $d=10$, 50, 100,
200 nm (from top to bottom) and (b) as a function of impurity location
$d$ for fixed impurity densities. 
Solid (dashed) lines represent  the percolation critical density,
$n_c^{per}$ (Ioffe-Regel critical density, $n_c^{IR}$). 
(c) and (d) show the results for p-GaAs with the same impurity
parameters of (a) and (b), respectively. 
(e) and (d) show $n_c$ for Si-MOSFET. In (e) the impurity locations
$d=1$, 10, 100 nm (from top to bottom) are used. 
\label{fig1}
}
\end{figure}

Several general comments can be made about the results shown in
Fig.~\ref{fig1}: (i) The analytically derived scaling behavior derived
earlier in this paper apply in their respective regimes of validity,
but the dependence of $n_c$ on both ($n_i$,
$d$) characterizing disorder precludes any definitive dependence of
$n_c$ on the system mobility since $n_c(n_i,d)$ and $\mu_m(n_i,d)$ at
some high density $n_m\gg n_c$ are two independent functions of $n_i$
and $d$. (ii) In general, $n_c^{IR} > n_c^{per}$ for larger values of
$n_i$ and/or $d$. We see the clear trend in Fig.~\ref{fig1} that as
$n_i$ ($d$) increases for fixed $d$ ($n_i$) respectively, $n_c^{IR}$
lines cross above the $n_c^{per}$ lines for all three 2D systems we
study. For lower disorder (i.e., smaller $n_i$), which is of
particular interest to 2D MIT phenomena, $n_c^{IR}$ always is smaller
than $n_c^{per}$. We expect the percolation theory to be of validity
only for rather large values of $d$ (since only then the Coulomb
disorder is effectively unscreened and leads to long-range potential
fluctuations in the 2D landscape), and again for `$d$' not too large,
we always find $n_c^{IR} < n_c^{per}$. (iii) For similar disorder
parameters (i.e., same values of $n_i$ and $d$), our results in
Fig.~\ref{fig1} indicate very similar (but not identical) values of
$n_c^{IR}$ for all three systems we study -- of course $n_c^{per} =
0.1 \sqrt{n_i}/d$ is, by definition, independent of the materials
parameters. This finding of similar $n_c^{IR}$ in all three systems,
while being surprising at first sight, turns out to be consistent with
experimental observations where the discrepancy in the reported $n_c$
values among different systems (with $n_c^{Si} \sim 10^{11} cm^{-2} >
n_c^{p-GaAs} \sim 10^{10} cm^{-2} > n_c^{n-GaAs} \sim 10^9 cm^{-2}$)
appears to arise almost entirely from the very different disorder
parameters in these systems (with $\mu_m^{Si} \sim 5 \times 10^4
cm^2/Vs < \mu_m^{p-GaAs} \sim 5\times 10^5 cm^2/Vs < \mu_m^{n-GaAs}
\sim 5 \times 10^6 cm^2/Vs$), where $\mu_m$ is the typical
high-density mobility value, more or less explain the difference in
their observed $n_c$ values based on the  
approximate scaling law $n_c \sim \mu_m^{-\gamma}$.
(iv) To the extent
the numerical results in Fig.~\ref{fig1} allow us to discern any
materials trend in the $n_c^{IR}$ values, we find that for the same
disorder strength (i.e., same values of $n_i$ and $d$) Si-MOSFETs tend
to have the lowest $n_c^{IR}$ with n-GaAs and p-GaAs having almost the
same calculated $n_c$, thus verifying the effective mass independence
of $n_c^{IR}$ we derived before. A clear prediction of this finding is
that 2D n-GaAs and 2D p-GaAs will have very similar values of $n_c$
provided they have similar disorder configurations. 

\begin{figure}[t]
	\centering
	\includegraphics[width=.9\columnwidth]{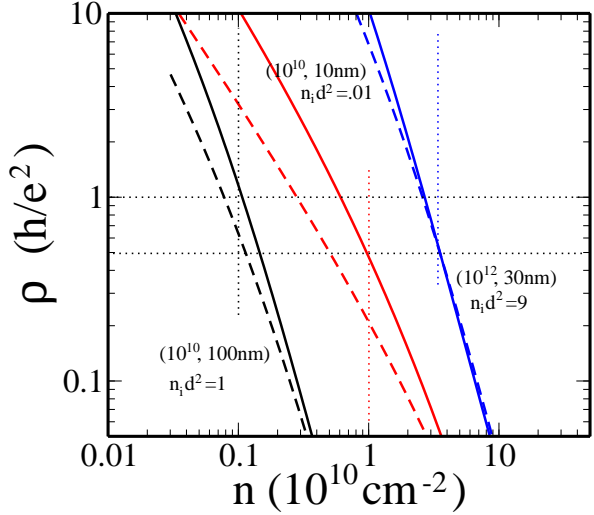}
	\caption{The calculated resistivity $\rho$ as a function of
          carrier density. The solid (dashed) lines show results for
          n-GaAs (Si-MOSFET). The sets of the
remote impurity density $n_i$ and the distance $d$ are ($10^{10}
cm^{-12}$, $d=100$nm) (i.e., $n_id^2 =1$); 
($n_i = 10^{10} cm^{-12}$, $d=10$nm) (i.e., $n_id^2 = 0.01$); ($n_i =
10^{12} cm^{-12}$, $d=30$nm) (i.e., $n_id^2 =9$) [from left to
  right]. The vertical dot lines indicate the percolation critical
density for given impurity conditions (i.e., $n_c = 0.1
\sqrt{n_i}/d$).  
\label{fig2}
}
\end{figure}

To reinforce the point that the Ioffe-Regel criterion typically leads
to $n_c$ values which depend strongly on the disorder, but only weakly
on the material, we show in Fig.~\ref{fig2} our calculated resistivity
$\rho(n)$ as a function of 2D carrier density $n$ for the n-Si-MOSFET
and the n-GaAs system for exactly the same set of values of ($n_i$,
$d$) with three different sets of disorder configurations (i.e., $n_i$
and $d$ values) shown in the plots. The $k_F l=1$ Ioffe-Regel
criterion translates into $\rho = h/e^2$, which gives similar $n_c$
values for the three sets of disorder shown in Fig.~\ref{fig2}. For
the purpose of comparison we also shown $n_c^{per} = 0.1
\sqrt{n_i}/d$, which again is reasonably  
close to the calculated $n_c^{IR}$ value for each disorder
configuration. At first sight, it appears that for the intermediate
disorder strength (red curves with $n_id^2 = 0.1$), the $n_c^{IR}$
values for Si and n-GaAs are very different from each other, but this
discrepancy is resolved once the valley degeneracy effect
(i.e. $g_v=2$ for Si) is taken into account so that the critical
resistivity $\rho_c = h/2e^2$ for the Si system.  
It becomes clear that if we use $\rho_c^{Si} = 0.5 h/e^2$ and
$\rho_c^{GaAs}=h/e^2$, then indeed the resultant $n_c$ values for the
two systems are very close to each other, indicating the approximate
materials universality of $n_c$ among different 2D systems, with
disorder being the primary determinant of $n_c$.

One   unexpected aspect of the results shown in Fig.~\ref{fig2} is
that the critical resistivity $\rho_c = \rho(n_c)$ for the percolation
transition seems to be not very different from that (i.e., $\rho_c
\sim h/e^2$) implied by the Ioffe-Regel criterion, which is, of
course, a direct manifestation of $n_c^{per}$ and $n_c^{IR}$ being not
that different (typically $n_c^{per} \agt n_c^{IR}$ in
Figs.~\ref{fig1} and \ref{fig2}) from each other. We show in
Figs.~\ref{fig3} -- \ref{fig6} our calculated $\rho_c^{IR}$ and
$\rho_c^{per}$ defined by 
\begin{eqnarray}
\rho_c^{IR} & = & \rho(n=n_c^{IR}), \nonumber \\
\rho_c^{per} & = & \rho(n=n_c^{per}).
\end{eqnarray}
Although we expect $\rho_c^{IR} \alt h/e^2$, by definition, there is
no reason for $\rho_c^{per}$ to have anything to do with $h/e^2$ since
it is a nonuniversal quantity not determined by quantum interference
or quantum localization. Our results, however, indicate that in
general $\rho_c^{per} \alt h/e^2$ as well!

\begin{figure}[t]
	\centering
	\includegraphics[width=1.\columnwidth]{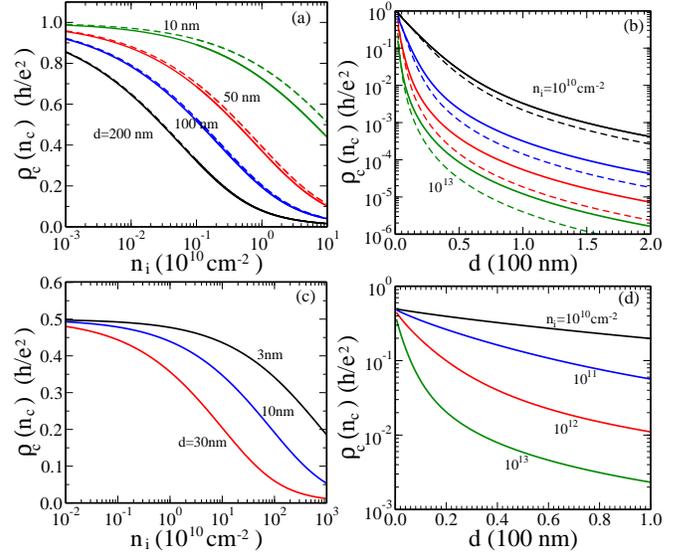}
	\caption{The calculated resistivity $\rho_c$ at Ioffe-Regel
          critical density, $n_c^{IR}$, (a) as a function of $n_i$ for
          different $d=10$, 50, 100, 200 nm and (b) $\rho_c$ as a
          function $d$ for different impurity densities $n_i=10^{10}$,
          $10^{11}$, $10^{12}$, $10^{13}$ cm$^{-2}$ (from top to
          bottom). The solid (dashed) lines are the results of n-GaAs
          (p-GaAs).  
(c) and (d) show the calculated resistivity $\rho_c$ for a n-Si-MOSFET
          (c) as a function of $n_i$ for different $d=3$, 10, 30 nm
          and (d) as a function $d$ for different impurity densities
          $n_i=10^{10}$, $10^{11}$, $10^{12}$, $10^{13}$ cm$^{-2}$. 
\label{fig3}
}
\end{figure}

In Fig.~\ref{fig3} we show our numerically calculated $\rho_c^{IR}$ as
a function of $n_i$ and $d$ for 2D n-GaAs, p-GaAs, and n-Si-MOS
systems. We emphasize that $\rho_c^{IR} = 2h/(g_sg_v e^2)$ universally
by definition if the quantity `$l$' in the $k_F l=1$ Ioffe-Regel
criterion is interpreted as the transport mean free path [see
  Eq.~(\ref{eq18})] of the 2D system. This means that $\rho_c^{IR} =
h/e^2$ (GaAs); $h/2e^2$ (Si) for all $n_i$ and $d$ if we take `$l$' to
be the transport mean free path $l=l_t$ as in Eq.~(\ref{IRc1}). All
our numerical $\rho_c^{IR}$ results therefore interpret $l=l_q$ as the
quantum mean free path [using Eq.~(\ref{IRc2}) without the
  conductivity vertex correction term] where $k_F l_q = 1$ becomes 
equivalent to $\Gamma = E_F$ strong localization condition. The most
important qualitative conclusion based on the numerical results of
Fig.~\ref{fig3} is that $\rho_c^{IR}$ is large (small) for small
(large) $n_i$ and small (large) $d$. In Fig.~\ref{fig3}
$\rho_c(n_i,d)$ falls off monotonically either as a function of
increasing $n_i$ or increasing $d$, which of course makes sense since
small $n_i$ and $d$ implies very small $n_c$, and hence rather large
$\rho_c^{IR}$ (which is still bounded from above by $h/e^2$ since
$\rho_c^{IR} \leq h/e^2$ by definition since $l_q \leq l_t$). The
decrease of $\rho_c^{IR}$ to incredibly small values as a function of
increasing $n_i$ or $d$ may appear completely unphysical (perhaps even
ridiculous) at first, but this is a direct manifestation of our model
of disorder which is entirely characterized  
by a 2D impurity plane containing $n_i$ random charged impurities per
unit area separated by a distance `$d$'. For large `$d$', this model
fails completely since there would always be some unknown and
unintentional background charged impurities which will cause the
strong localization crossover at some higher value of $\rho_c$ (i.e.,
lower value of $n_c$). In principle, however, the qualitative result
emerging from Fig.~\ref{fig3} is that more disordered the system
(i.e., larger the value of $n_i$), lower is the critical resistance
$\rho_c$ at the transition (and higher is the $n_c$). This is
certainly qualitatively correct since older MOSFETs (before 1994 -- 95
when the current era of 2D MIT physics commerced with the Kravchenko
{\it et al.} work \cite{kravchenko1995}) typically had \cite{andormp}
very high values of $n_c$ 
($> 10^{12}$ cm$^{-2}$) with consequently rather low values of
$\rho_c$ ($\sim h/10e^2 \approx 2$ k$\Omega$) \cite{andormp}. We also
mention in this context the empirical finding of Sarachik
\cite{sarachikepl} that $n_c \sim \mu_m^{-0.67}$ which implies very
large $n_c$ (and hence rather low $\rho_c$) for samples with very
large values of $n_i$ (i.e., low values of $\mu_m$). 
Results with very large $d$-values in
Fig.~\ref{fig3} are shown only for the sake of completeness since
other unknown disorder with small `$d$' (not included in the model)
will intervene making our large $d$ results inapplicable to
experimental systems. Our results, however, do indicate that extremely
pure modulation doped 2D samples with large values of `$d$' should
have relatively small values of $\rho_c$ if all other disorder effects
are absent. 
This theoretical prediction should be experimentally tested in the
future.

\begin{figure}[t]
	\centering
	\includegraphics[width=1.\columnwidth]{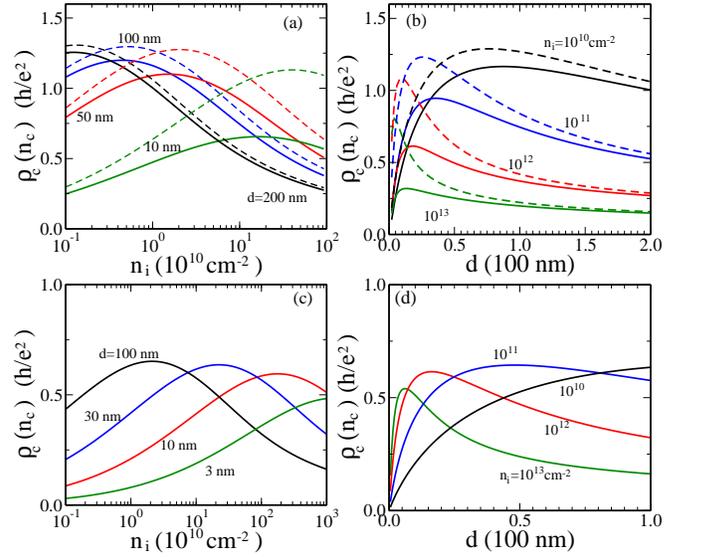}
	\caption{The calculated resistivity $\rho_c$ at percolation
          critical density, $n_c^{per}$, (a) as a function of $n_i$
          for different $d=10$, 50, 100, 200 nm and (b) as a function
          $d$ for different impurity densities $n_i=10^{10}$,
          $10^{11}$, $10^{12}$, $10^{13}$ cm$^{-2}$. The solid
          (dashed) lines are the results of n-GaAs (p-GaAs).  
(c) and (d) show the calculated resistivity $\rho_c$ for a n-Si-MOSFET
          (c) as a function of $n_i$ for different $d=3$, 10, 30, 100
          nm and (d) as a function $d$ for different impurity
          densities $n_i=10^{10}$, $10^{11}$, $10^{12}$, $10^{13}$
          cm$^{-2}$. 
\label{fig4}
}
\end{figure}

In Fig.~\ref{fig4} we show the same results as in Fig.~\ref{fig3}
except now for the percolation theory (i.e. $\rho_c^{per}$ is shown as
a function of $n_i$ and $d$ in Fig.~\ref{fig4} in contrast to
Fig.~\ref{fig3} where $\rho_c^{IR}$ is shown).  It is clear that
$\rho_c^{per}$ (Fig.~\ref{fig4}) behaves qualitatively very
differently than $\rho_c^{IR}$ (Fig.~\ref{fig3}) with $\rho_c^{per}
\sim h/e^2$ within a factor of 2 for most values of $n_i$ and $d$. (We
emphasize again that $\rho_c^{IR} = h/e^2$ within a factor of 2 also
if the Ioffe-Regel criterion is taken to be $l=l_t$ in the $k_F l=1$
condition.) The fact that the percolation transition which defines
$n_c^{per} = 0.1 \sqrt{n_i}/d$ with $\rho_c^{per} = \rho_c(n_c)$ as
obtained from our standard Drude-Boltzmann semiclassical transport
theory provides a very 
reasonable value of $\rho_c \sim h/e^2$ for a wide range of realistic
disorder parameters is certainly somewhat of a surprise. We should
mention that for unrealistically large $n_i$ and/or unrealistically
small $d$, $\rho_c^{per}$ takes on unrealistic values, but for
realistic physical combinations of ($n_i$, $d$) values operational in
real 2D systems our theoretical $\rho_c^{per}$ seems to agree well
with the experimental results. We do not know at this stage whether
this is simply a coincidence or indicates some deep truth about the
importance of percolation transport in the 2D MIT phenomena. 

\begin{figure}[t]
	\centering
	\includegraphics[width=1.\columnwidth]{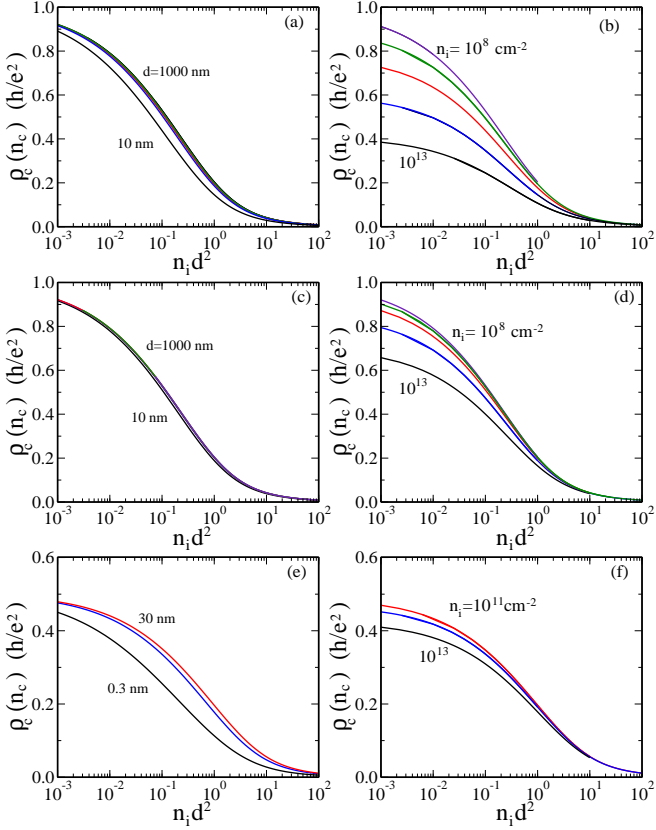}
	\caption{(a) and (b) show the calculated resistivity at
          Ioffe-Regel critical density $n_c^{IR}$, $\rho_c^{IR}$, for
          n-GaAs as a function of $n_id^2$ (a) varying $n_i$ for
          various fixed $d=10$, 50, 100, 200, 1000 nm (from bottom to
          top) and (b) varying $d$ for various fixed $n_i=10^8$,
          $10^{10}$, $10^{11}$, $10^{12}$, $10^{13}$ cm$^{-2}$ (from
          top to bottom), respectively.  
(c) and (d) show $\rho_c^{IR}$ for p-GaAs with the same impurity
          parameters used in (a) and (b), respectively.  
(e) and (f) show $\rho_c^{IR}$ as a function of $n_id^2$ for Si-MOSFET
(e) varying $n_i$ for various fixed $d=0.3$, 3, 30 nm (from bottom to
          top) and (f) varying $d$ for various fixed $n_i=10^{11}$,
          $10^{12}$, $10^{13}$ cm$^{-2}$ (from top to bottom),
          respectively.  
\label{fig5}
}
\end{figure}

Results shown in Figs.~\ref{fig3} and \ref{fig4} hint at the
dimensionless parameter $n_i d^2$ being the important disorder
parameter determining $\rho_c^{IR}$ and $\rho_c^{per}$. This, in fact,
follows from the definitions of these two critical
resistivities. Using the Boltzmann transport theory for charged
impurity scattering limited transport at $T=0$, \cite{dassarmaprb2013}
we find 
\begin{equation}
\rho_c = \frac{8h}{e^2} \frac{n_i}{n_c} x_c^2 \int_0^1dy
\frac{dy}{\sqrt{1-y^2}} \frac{y^2 e^{-2yd_c}}{(y+x_c)^2}, 
\label{eq54}
\end{equation}
where
\begin{equation}
k_c = k_F(n_c) = \left ( \frac{4\pi n_c}{g_sg_v} \right )^{1/2}; \;
x_c = q_{TF}/2k_c; \; d_c = 2k_c d. 
\end{equation}
Putting $n_c^{per}$ or $n_c^{IR}$ for $n_c$ in Eq.~(\ref{eq54}) we
obtain $\rho_c^{per}$ and $\rho_c^{IR}$ respectively. We note that the
explicit dependence of $\rho_c \sim n_i$ in Eq.~(\ref{eq54}) is
misleading since $n_c$ itself has an $n_i$ dependence also.

\begin{figure}[t]
	\centering
	\includegraphics[width=1.\columnwidth]{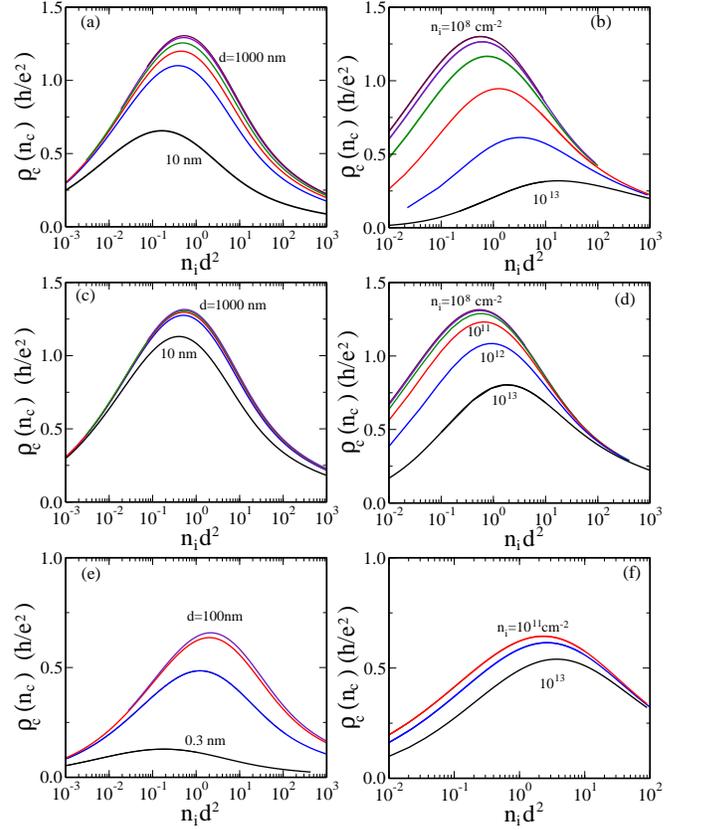}
	\caption{ (a) and (b) show the calculated resistivity at
          percolation critical density $n_c^{per}$, $\rho_c^{per}$,
          for n-GaAs as a function of $n_id^2$ (a) varying $n_i$ for
          various fixed $d=10$, 50, 100, 200, 1000 nm (from bottom to
          top) and (b) varying $d$ for various fixed $n_i=10^8$,
          $10^{10}$, $10^{11}$, $10^{12}$, $10^{13}$ cm$^{-2}$ (from
          top to bottom), respectively.  
(c) and (d) show $\rho_c^{per}$ for p-GaAs with the same impurity
          parameters used in (a) and (b), respectively.  
(e) and (f) show $\rho_c^{per}$ as a function of $n_id^2$ for Si-MOSFET
(e) varying $n_i$ for various fixed $d=0.3$, 3, 30, 100 nm (from
          bottom to top) and (f) varying $d$ for various fixed
          $n_i=10^{11}$, $10^{12}$, $10^{13}$ cm$^{-2}$ (from top to
          bottom), respectively. 
\label{fig6}
}
\end{figure}

The integral on the right hand side of Eq.~(\ref{eq54}) can be
analytically evaluated in various asymptotic limits for both
$\rho_c^{IR}$ and $\rho_c^{per}$, giving the following results (with
$k_c \sim \sqrt{n_c}$): For $k_cd \gg1$ 
\begin{eqnarray}
\rho_c^{IR} & \propto & (n_i d^2)^{-2/3}; \;\; n_c \sim (n_i/d)^{2/3},
\nonumber \\ 
\rho_c^{per} &  \propto & (n_id^2)^{-1/4}; \;\; n_c \propto \sqrt{n_i}/d.
\label{eq56}
\end{eqnarray}
For $k_c d \ll 1$
\begin{eqnarray}
\rho_c^{IR} & \propto & (n_i d^2)^{-3/2}; \;\; n_c \sim n_id^0, \nonumber \\
\rho_c^{per} &  \propto & (n_id^2)^{1/2}; \;\; n_c \propto \sqrt{n_i}/d.
\label{eq57}
\end{eqnarray}
This shows that $n_id^2$ is an important dimensionless parameter
determining the disorder scaling of the crossover resistivity
$\rho_c$. We have explicitly checked numerically that these equations
[Eqs.~(\ref{eq56}) and (\ref{eq57})] are in quantitative agreement of
our numerical results.

We note that $\rho_c$ and $n_c$ have very different qualitative
dependence on the disorder parameters $n_i$ and $d$, and this might
enable an experimental  
distinction between them possible if quantitative information about
the underlying disorder becomes available. In Figs.~\ref{fig5} --
\ref{fig7} we show our numerically calculated $\rho_c$ as a function
of $n_id^2$ to explicitly depict the dimensionless dependence of
$\rho_c/(h/e^2)$ on the dimensionless disorder parameter $n_i d^2$. 

\begin{figure}[t]
	\centering
	\includegraphics[width=1.\columnwidth]{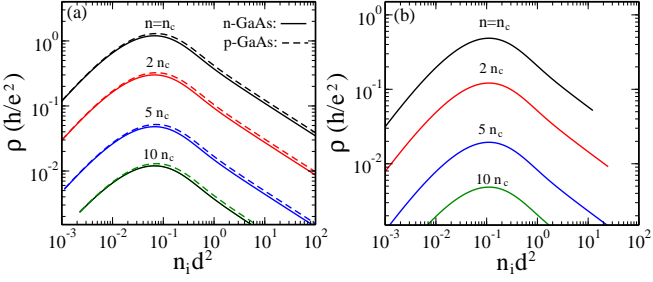}
	\caption{The calculated resistivity $\rho_c^{per}$ as a
          function of $n_id^2$ (a) for n- and (solid lines) p-GaAs
          (dashed lines), and (b) for Si-MOSFET. In (a) and (b)
          $d=100$ nm and  $d=3$ nm are used respectively. The
          percolation critical density $n_c=0.1\sqrt{n_i}/d$ is
          calculated by changing $n_i$. The resistivity $\rho$ is
          calculated at different values of $n \geq n_c^{per}$ in
          order to show the scaling behavior. 
\label{fig7}
}
\end{figure}

In Figs.~\ref{fig5} and \ref{fig6} we show our calculated
$\rho_c^{IR}$ and $\rho_c^{per}$ as a function of $n_i d^2$ for
various fixed values of $n_i$ and $d$ (as shown in the figures) for 2D
n-GaAs, p-GaAs, and n-Si-MOSFET systems. It is clear (which is also
obvious from our analytical results) that the Ioffe-Regel and the
percolation criteria provide very different qualitative dependence of
$\rho_c$ on disorder parameters. Finally, in Fig.~\ref{fig7} we show
the calculated resistivity $\rho$ at different values of $n \geq
n_c^{per}$ in order to emphasize the scaling behavior.

\subsubsection{Realistic 2D structures}

All results shown in Figs.~\ref{fig1} -- \ref{fig7} are for strict
zero-thickness 2D
systems with the appropriate effective mass, lattice dielectric
constant, and valley degeneracy ($g_v = 1$, 2 for GaAs, Si,
respectively) defining each semiconductor material. Results given in
Figs.~\ref{fig1} -- \ref{fig7} serve to provide the qualitative
dependence of the critical density and resistivity on disorder
parameters, but are not quantitatively realistic even if the disorder
parameters (i.e., $n_i$ and $d$) were precisely known. In particular,
the finite quantum thickness of the realistic quasi-2D system softens
the Coulomb disorder arising from the charged impurities since the 2D
Coulomb interaction changes from $2\pi e^2 /\kappa q$ to $(2\pi
e^2/\kappa q) f(q)$ where $f(q) \leq 1$ is the quasi-2D form factor
due to the finite quantum thickness effect [and $f(q)=1$ in the ideal
  2D limit]. 
Since the modification to the transport theory for $f(q) < 1$ is
well-known \cite{andormp,sarmaprb2004} we do not provide any details,
concentrating instead on the numerical results for $n_c$ in the
realistic quasi-2D situation.

\begin{figure}[t]
	\centering
	\includegraphics[width=1.\columnwidth]{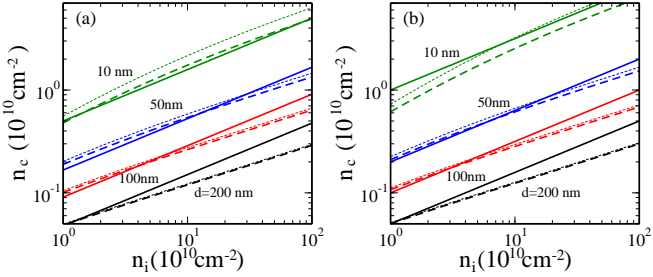}
	\caption{(a) Calculated $n_c^{IR}$ (dashed lines) and
          $n_c^{per}$ (solid lines)  as a function of impurity density
          for  fixed several $d=10$, 50, 100, 200 nm  
	for GaAs quantum wells with a well width $a=200$ \AA. Thick
        (thin) dashed lines represent results for n-GaAs (p-GaAs). (b)
        The same results as (a) for zero quantum well thickness (i.e.,
        $a=0$). 
\label{fig8}
}
\end{figure}

\begin{figure}[t]
	\centering
	\includegraphics[width=.8\columnwidth]{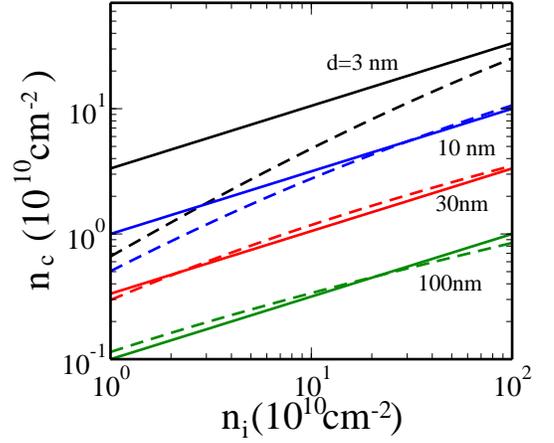}
	\caption{The calculated critical density of n-MOSFET as a
          function of impurity density $n_i$ for fixed impurity
          locations $d=3$, 10, 30, 100 nm (from top to bottom). Here
          the the finite thickness of quasi-2D system is considered.  
Solid (dashed) lines represent  the percolation critical density,
$n_c^{per}$ (Ioffe-Regel critical density, $n_c^{IR}$). 
\label{fig9}
}
\end{figure}

In Fig.~\ref{fig8} we show our $n_c^{IR}$ and $n_c^{per}$ results
[Fig.~\ref{fig8}(a)] for n- and p-GaAs quantum wells (using $\tau =
\tau_t$ so that $\rho_c^{IR} = h/e^2$). For the purpose of comparison,
we also provide our results for the strict 2D limit (i.e., zero
quantum well thickness $a=0$) in Fig.~\ref{fig8}(b). For $n_c^{per} =
0.1\sqrt{n_i}/d$, the only effect of finite well-thickness ($a\neq 0$)
is that the effective value of `$d$' changes by $a$, changing
$n_c^{per}$ to $n_c^{per} = 0.1 \sqrt{n_i}/(d+a/2)$. For $n_c^{IR}$,
the finite thickness increases the effective mean free path $l$, and
thus suppress the resultant $n_c$. Thus, 
both $n_c^{IR}$ and $n_c^{per}$ are suppressed by the finite thickness
with this suppression effect being very strong for $n_c^{per}$ when
$d<a$. A comparison of Figs.~\ref{fig8}(a) and (b) bear this out, and
thus the finite thickness effect is only quantitative with the
qualitative power law dependence of $n_c$ on $n_i$ being approximately
the same.

In Fig.~\ref{fig9} we show our realistic quasi-2D results for
n-Si-MOSFETs where the quasi-2D quantum thickness is determined
self-consistently by the carrier density $n$ itself \cite{andormp}
which we incorporate through the variational Stern-Howard wavefunction
\cite{stern}. We note that for small values of $d$ (which is the
expected situation in Si-MOSFETs since the charged impurities are
typically in the SiO$_2$ layer close to the Si-SiO$_2$ interface),
$n_c^{IR} \sim 10^{10} - 10^{11}$ cm$^{-2}$ for $n_i \sim
10^{10}-10^{11}$ cm$^{-2}$ whereas $n_c^{IR} > 10^{11}$ cm$^{-2}$ for
$n_i > 10^{11}$ cm$^{-2}$. These finding are consistent with the
higher- and lower-mobility Si-MOSFET devices, respectively.

\begin{figure}[t]
	\centering
	\includegraphics[width=1.\columnwidth]{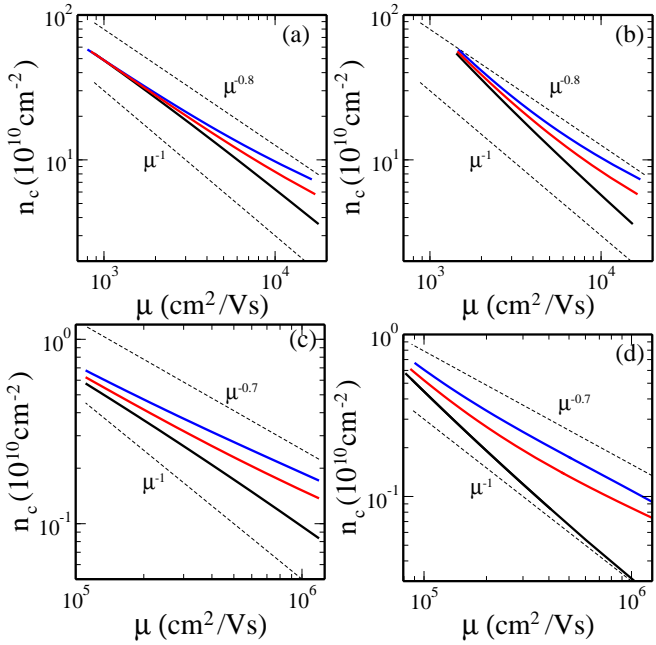}
	\caption{(a) and (b) show the  calculated Ioffe-Regel critical
          density $n_c^{IR}$ of Si-MOSFET as a function of mobility
          for different temperatures, $T=0$, 1, and 2K (from bottom to
          top). The mobility is calculated (a) at a given high density
          $n = 5 \times 10^{11}$ cm$^{-2}$ and (b) at $n=5n_c$. 
(c) and (d) show the $n_c$ of n-GaAs as a function of mobility for
          different temperatures, $T=0$, 0.2, and 0.5K (from bottom to
          top). The mobility is calculated (c) at a given high density
          $n = 5 \times 10^{10}$ cm$^{-2}$ and (b) at $n=5n_c$. 
\label{fig10}
}
\end{figure}

One important qualitative point to note in Figs.~\ref{fig8} and
\ref{fig9} is that while there is a large difference between
percolation and Ioffe-Regel predictions for $n_c$ for large values of
`$d$', for small values of $d$, they are virtually
indistinguishable. We also note that the materials difference (e.g.,
n- versus p-GaAs 2D systems in Fig.~\ref{fig8}) is rather small with
respect to the calculated $n_c$ for the same disorder. 
It may be worthwhile to point out that writing $n_c \sim n_i^{\delta}$
in Fig.~\ref{fig9}, we get $\delta = \delta(d,n_i)$, and our best
numerical estimate for the exponent $\delta$ is: $\delta \approx
0.8-1$ for $d=1$ nm, $\delta \approx 0.6-0.9$ for $d=5$ nm, and
$\delta \approx 0.5-0.8$ for $d=15$ nm. Since $\mu \sim n_i^{-1}$, we
can approximate $\gamma = \delta$ (where $n_c \sim \mu_m^{-\gamma}$),
and thus our earlier estimate of $\gamma \approx 0.67$ for Si-MOSFET
is consistent with $d=1-2$ nm. 
This is a stringent consistency check on our theory since, indeed, the
random charged impurities in Si MOSFETs are known to be located $1-2$
nm inside the oxide layer near the Si-SiO$_2$ interface.

In Fig.~\ref{fig10} we show our calculated Ioffe-Regel value of $n_c$
as a function of a fiduciary ``maximum mobility" defined as the
mobility calculated for exactly the same value of disorder parameters
(i.e., the same sample), but at a much higher density $n_m \gg n_c$. 
The precise dependence of $n_c$ on the high-density mobility $\mu$, of
course, depends somewhat on the fiduciary density chosen for the
high-density mobility, but the basic finding is that the power law
($\gamma$) dependence, $n_c \sim \mu_m^{-\gamma}$, is a function of
temperature, and typically $\gamma \sim 0.7-0.8$ as already pointed
out empirically by Sarachik a long time ago \cite{sarachikepl}. The
fact that $\gamma \approx 0.6-0.8$ is consistent with experimental
findings in different systems is an indication that the experimental
2D MIT is likely to be a strong localization crossover phenomenon. 
One salient feature of the results presented in Fig.~\ref{fig10} is
that the effective exponent $\gamma$, 
$n_c \sim \mu_m^{-\gamma}$ where $\mu_m$ is the mobility at same high
density $n_m \gg n_c$, depends strongly on the temperature (as one
would expect because of the strong temperature dependence of the 2D
metallic resistivity for $n \agt n_c$ provided $n_c$ is not too
large).

\begin{figure}[t]
	\centering
	\includegraphics[width=1.\columnwidth]{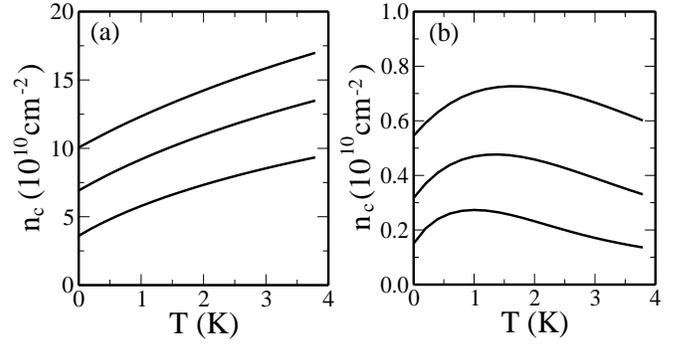}
	\caption{Calculated Ioffe-Regel $n_c(T) $ of (a)  Si-MOSFET as
          a function of temperature for different impurity densities
          $n_i =0.5$, 1.0, $1.5\times 10^{11}$ cm$^{-2}$ (from bottom
          to top) and (b) n-GaAs for $n_i = 0.2$, 0.5,
          $1.0\times10^{10}$ cm$^{-2}$ (from bottom to top). 
\label{fig11}
}
\end{figure}

In Fig.~\ref{fig11} we show our numerically calculated
$n_c(T)$, based on the finite-temperature Ioffe-Regel criterion $k_F
l(T) = 1$, as a function of temperature. As discussed earlier,
$n_c(T)$ first increases with $T$ and then decreases when $T \alt
T_F$. However, the overall variation in $n_c(T)$ is less than a factor
of 2 in our results. 
We mention that the results shown in Fig.~\ref{fig11}(a) and (b) agree well respectively with the experimentally measured temperature dependence of the critical 2D MIT density in Si MOSFETs \cite{tracyprb2009} and 2D  electrons \cite{lillyprl2003} and holes \cite{manfraprl2007} in GaAs systems, providing strong support for our basic Ioffe-Regel model describing the 2D MIT crossover. 

\begin{figure}[t]
	\centering
	\includegraphics[width=1.\columnwidth]{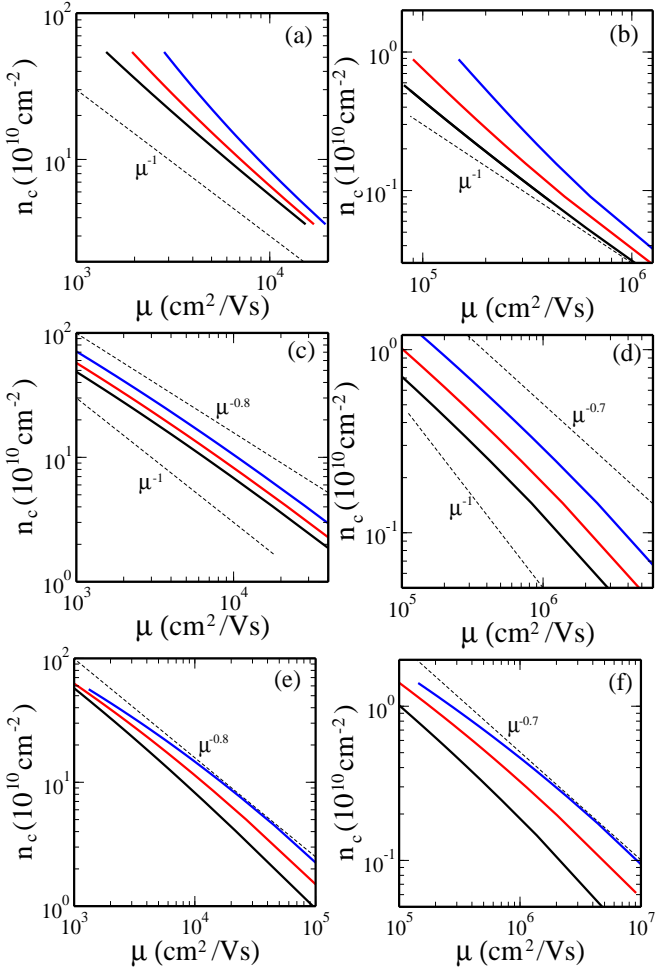}
	\caption{Calculated zero temperature Ioffe-Regel critical
          densities (solid lines) as a function of a reference
          mobility for (a) Si-MOSFET and (b) for n-GaAs. The reference
          mobility is calculated for different densities $n = 5n_c$,
          $10n_c$, $20n_c$ (from bottom to top). 
Zero temperature $n_c$ as a function of mobility calculated at (c) $n
= 5$, 10, $20 \times 10^{11}$ cm$^{-2}$ (from bottom to top) for
Si-MOSFET and (d) $n = 5$, 10, 
$20 \times 10^{10}$ cm$^{-2}$ for n-GaAs (from bottom to top).
(e) Zero temperature $n_c$ of Si-MOSFET as a function of a reference
mobility calculated at $n = 10 \times 10^{11}$ cm$^{-2}$ for different
locations of impurity center, $d = 0$, 10, and 20 \AA \; (from bottom
to top), and (f) $n_c$ of n-GaAs as a function of a mobility at $n =
10 \times 10^{10}$ cm$^{-2}$ for different locations of impurity
center, $d = 0$, 50, and 100 \AA \; (from bottom to top). 
	\label{fig12}
}
\end{figure}

\begin{figure}[t]
	\centering
	\includegraphics[width=1.\columnwidth]{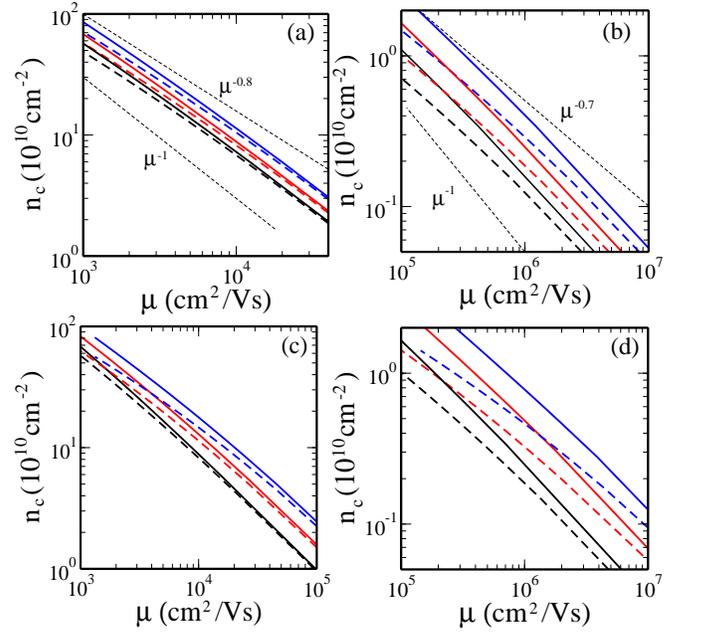}
	\caption{(a) and (b) show, respectively,  the calculated $n_c$
          with the quantum mean free path using the same parameters of
          Fig.~\ref{fig12}(c) and (d) where the transport mean free
          path is used. Solid (dashed) lines indicate the results with
          quantum (transport) mean free path. 
(c) and (d) show the calculated $n_c$ with the quantum mean free path
          using the same parameters of Fig.~\ref{fig12}(e) and (f)
          where the transport mean free path is used,
          respectively. Solid (dashed) lines indicate the results with
          quantum (transport) mean free path. 
	\label{fig13}
}
\end{figure}

In Fig.~\ref{fig12}, we show that the calculated maximum mobility
dependence of $n_c$ is to some  
extent dependent on how the maximum mobility is chosen, and thus one
cannot really discuss a unique dependence of $n_c$ on the maximum
mobility, which is obvious from the fact that both $n_c = n_c(n_i,d)$
and the mobility $\mu = \mu(n_i,d)$ are independent functions of $n_i$
and $d$. What is interesting, however, is the finding that the
exponent $\gamma$ (with $n_c \sim \mu_m^{-\gamma}$) remains within our
analytical finding of $\gamma \approx 0.6 - 0.8$ for a wide range of
definitions of the maximum mobility $\mu$.

In Fig.~\ref{fig13} we show how our realistic numerical results change
if the quantum mean free path with $l=l_q=v_F \tau_q$ is used in the
$k_F l=1$ criterion for 2D MIT. There is no qualitative change in the
results with $\tau_q$ replacing $\tau_t$ in the Ioffe-Regel criterion
as we already emphasized earlier in this paper.

Finally, in Figs~\ref{fig14} and \ref{fig15} we show the effect of an
applied parallel magnetic field $B$ on the critical density $n_c(B)$
due to the spin-polarization-induced lifting of spin degeneracy $g_s$
from $g_s=2$ at $B=0$ to $g_s=1$ at $B=B_s$ where $B_s$ is the applied
field strength to fully spin-polarized the 2D electrons. We show
numerical results only for Si-MOSFETs here since the qualitative
effect of the parallel field on 2D MIT is the same for all 2D systems
since the relevant physics is the suppression of screening (and hence
suppression of the transport mean free path $l$) due to the applied
magnetic field. We neglect all orbital effects 
of the applied magnetic field which could enhance $n_c(B)$ even more
for systems with large quasi-2D thickness \cite{sarmaprl2000}. As
mentioned already, the maximum possible effect of the magnetic field
is an enhancement of $n_c$ by a factor of $\sqrt{2}$ due to the
reduction of spin degeneracy from 2 to 1. Thus, our results in
Fig.~\ref{fig14} show an approximate 40\% enhancement of $n_c$ in the
presence of the applied field at $T=0$ whereas at finite temperatures
the effect is smaller. We emphasize that although the
spin-polarization-induced enhancement of $n_c(B)$ compared with its
$B=0$ value is a universal qualitative phenomenon as long as screened
Coulomb disorder is the dominant underlying transport scattering
mechanism, 
the actual quantitative effect would be miniscule (and experimentally
unobservable) if $n_c(B=0) = n_c$ is very large (as it is highly
disordered 2D systems where the 2D MIT phenomena have no dramatic
consequences) since $B_s = 2 E_c/g\mu_B$ with $E_c = E_F(n=n_c)$ would
be very large when $n_c$ is large, and thus $B/B_s \ll 1$ limit would
apply on any physically applicable magnetic field in the laboratory
making $n_c(B) \approx n_c(B=0)$. It is only when $n_c$ [and hence
  $E_c = E_F(n_c)$] is sufficiently small that the  applied parallel
field induced enhancement of $n_c$ can be experimentally relevant
since the available laboratory applied field could reach the $B/B_s
\sim1$ regime. Thus, the condition for the observation of strong
temperature dependence of the metallic resistivity and the condition
for the observation of strong magnetic field dependence of 2D MIT are
closely related as they both require fairly small $n_c$ (and therefore
very high-quality 2D samples) so that $T/T_F$ and $B/B_s$ can be
relatively large in respective cases. This close connection between
the temperature dependence and the magnetic field dependence of 2D MIT
phenomena is experimentally well-established, and has already been
noted in the literature \cite{hwangmagnetic}.

\begin{figure}[t]
	\centering
	\includegraphics[width=1.\columnwidth]{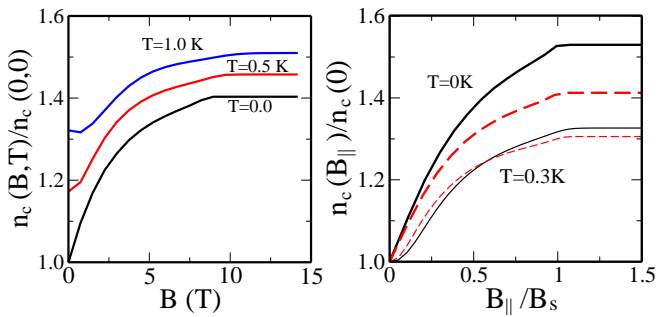}
	\caption{(a) Calculated $n_c(B)$ as a function of parallel
          magnetic field $B$ for 
an impurity density $n_i= 10^{11}cm^{-2}$ and for different
temperatures $T=0.0$, 0.5, 0.1 K. 
The critical density is normalized by the critical density calculated
at $B=0$ and $T=0$, where $n_c(0,0)=7\times 10^{10}cm^{-2}$.
Above $B_s$ the scattering time $\tau$ is constant in this model,
which gives the saturation of the critical density.
(b) Calculated $n_c(B)$ as a function of parallel magnetic field $B_{\|}$ for
two different impurity densities [$n_i=5\times 10^{10}$ (solid lines)
  and $10^{11}$ cm$^{-2}$ (dashed lines)] and for two temperatures
[$T=0K$ (thick lines) and $T=0.3K$ (thin likes)]. The 
critical density is normalized by the critical density calculated at
$B=0$ and at given temperature, $n_c(0,T)$.
\label{fig14}
}
\end{figure}

We note that at very low applied field values in Fig.~\ref{fig14},
there is a small upturn in the critical density compared with its
zero-field value.  This is a real effect arising from the increase in
effective $k_F$ induced by the applied field which always suppresses
$n_c$ at finite field compared with its zero-field value, as noted
earlier in this paper.  If screening effects are unimportant
(e.g. scattering by unscreened short-range disorder or at very high
carrier density with $2k_F \gg q_{TF}$), then this Fermi surface
effect would dominate the finite field transport properties.  But the
2D MIT phenomenon occurs at low values of $n_c$, where $q_{TF} >
2k_F$, and screening effects dominate, leading to a suppression of the
metallic phase and an increase of $n_c$ at finite applied magnetic
field.

Finally, in Fig.~\ref{fig15} we show our numerical results on the
valley-degeneracy dependence of $n_c$ by plotting the numerically
calculated $n_c(g_v)$ as a function of the valley degeneracy $g_v$ (at
fixed $g_s=2$) which we assume for this purpose to be a fictitious
continuous variable -- in reality $g_v$ is 1 or 2 for Si(100)-MOSFETs
[whereas for Si(111)-MOSFETs, $g_v=6$ is allowed]. As expected
$n_c(g_v)$ behaves very similarly to the spin-polarization effect on
$n_c$, and with decreasing valley degeneracy, $n_c$ is enhanced since
screening is reduced. This dependence of $n_c$ on spin and valley
degeneracy of the 2D system is consistent with detailed experimental
results reported in 2D AlAs system \cite{shayegannp}.  
We emphasize that our calculated $n_c(g_s)$ for fixed $g_v$ is
identical to results shown in Fig.~\ref{fig15} since both $g_s$ and
$g_v$ enter the theory equivalently as the product $g_sg_v$ through
the density of states. We mention that for $T\neq 0$ (not shown in
Fig.~\ref{fig15}), $n_c(g_v)$ softens somewhat showing a weaker
dependence of $n_c$ on $g_v$. The numerical $n_c(g_v)$ in
Fig.~\ref{fig15} agrees exactly with the analytical dependence
$n_c^{IR} \propto g_v^{-1}$ for $k_F d \ll 1$ which is satisfied
essentially for all values of $n_c$ shown in Fig.~\ref{fig15}. 

\begin{figure}[t]
	\centering
	\includegraphics[width=.8\columnwidth]{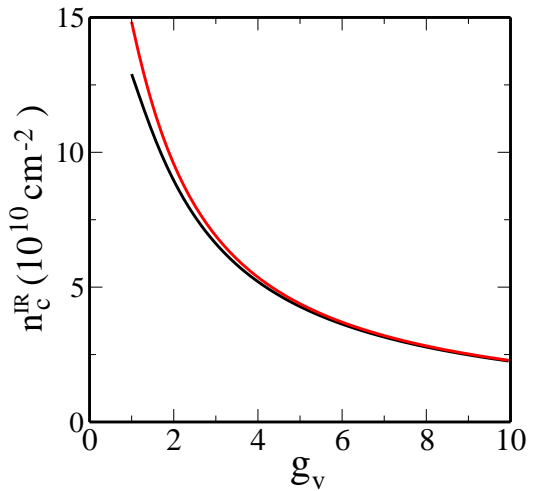}
	\caption{The calculated Ioffe-Regel critical density of
          Si-MOSFET as a function of valley degeneracy. Black solid
          (red dashed)
          line indicates the $n_c^{IR}$ calculated from the transport
          (quantum) scattering time $\tau_t$ ($\tau_q$).  
The results are shown for an ideal 2D system with zero thickness and
$d=1$ nm -- a finite thickness corresponds to a larger value of
effective `$d$', decreasing $n_c$ accordingly. 
\label{fig15}
}
\end{figure}

\section{discussion}

We have studied 2D MIT as a strong localization induced crossover
phenomenon determined by the Ioffe-Regel criterion 
(comparing with the corresponding classical percolation transition in the disorder-induced
`mountains-and-lakes' inhomogeneous potential fluctuations
landscape). There are several distinct aspects of the 2D MIT
phenomenology we have addressed in this work theoretically. 
The main results obtained in this paper involve theoretical and
numerical calculations of the crossover critical density $n_c^{IR}$
for 2D MIT using the Ioffe-Regel criterion and its comparison with the
corresponding percolation transition density $n_c^{per}$.

In addition to obtaining $n_c$, we also provide results for the
critical resistivity $\rho_c = \rho(n=n_c)$, which at $T=0$, is by
definition $\rho_c = 2h/(g_sg_ve^2)$. For a percolation transition at
$n=n_c^{per} = 0.1 \sqrt{n_i}/d$, in principle, $\rho_c^{per}$ could
have any value, but in practice $\rho_c^{per} \sim h/e^2$ seems to
apply extensively for realistic 2D sample parameters.

All our theoretical results use a minimal model of Coulomb
disorder characterized by a random 2D charged impurity density $n_i$
and a separation of $d$ between the impurities and the 2D
carriers. More complex models of disorder are straightforward to
include in the theory, but will involve more (than two) free
parameters, making it difficult to interpret and understand the
theoretical results. Since precise information about the details of
disorder is not typically available for high-quality 2D semiconductor
systems manifesting 2D MIT, our 2-parameter impurity model is a
reasonable starting point for discussing the 2D MIT phenomena.

Below we summarize and critically discuss our important findings and
related questions focusing on the key features of our theory  
as compared with the experimental 2D MIT phenomenology.

\subsection{Localization versus percolation}

A question of great importance, of course, is whether the 2D MIT at
$n\approx n_c$ is a strong localization quantum crossover or
classical percolation, i.e., whether $n_c = n_c^{IR}$ or
$n_c=n_c^{per}$. At first sight, it appears that our theory should be
able to answer this question with sharp precision since the physical
origin and the mathematical description of the Ioffe-Regel quantum
crossover and the percolation transition are completely distinct. This
turns out to be a much more difficult issue than anticipated at first
because $n_c^{IR}$ and $n_c^{per}$ have similar magnitudes (both in
qualitative agreement with experimental $n_c$) in many situations for
realistic values of sample parameters. This is a rather surprising
finding of our work that could not have been anticipated earlier. Even
more surprisingly we find that the critical resistivity $\rho_c =
\rho(n=n_c)$ is similar for both Ioffe-Regel theory and percolation
theory! This is a very unexpected and rather strange result since in
principle $\rho_c^{per} = \rho(n=n_c^{per}$) is allowed to be
arbitrary whereas $\rho_c^{IR}$ is closely related to the resistance
quantum $h/e^2 \approx 25,600 \Omega$ since it arises from quantum
localization. But our explicit calculations for 2D n-Si, n-GaAs, and
p-GaAs systems show that for realistic experimental parameters (with
$n_c\sim 10^9 - 10^{11}$ cm$^{-2}$), $\rho_c$ values calculated from
localization and percolation considerations are not widely different
although their dependences on system parameters could be quite
different. In view of this similarity between absolute values of $n_c$
(and $\rho_c$) in the two theories, it is not easy to manifestly
choose one mechanism over the other as determining the 2D MIT
crossover density $n_c$, at least using the
experimental data on $n_c$ and $\rho_c$ only.

One practical possibility is that as the carrier density is lowered
from the high-density metallic phase ($n \gg n_c$) to the low-density
insulating phase ($n<n_c$), whichever transition occurs first (i.e.,
at higher carrier density) in a given sample dominates the actual
crossover behavior in that system. (We mention here that our
theoretical calculation of $n_c$ and $\rho_c$ explicitly approaches
the transition from above, i.e., from the metallic phase.) Thus, $n_c
= n_c^{per}$ if $n_c^{per} > n_c^{IR}$ and $n_c = n_c^{IR}$ if
$n_c^{IR} > n_c^{per}$. At very low temperatures, however, quantum
interference must always be present and therefore $n_c \rightarrow
n_c^{IR}$ as $T \rightarrow 0$. Careful  
experiments should be carried out to investigate this question of
whether $n_c$ is better described as localization or as
percolation. We emphasize that the experimental finding that $\rho_c
\sim h/e^2$ (typically within a factor of $2-3$) does not
automatically imply that $n_c$ is described by $n_c^{IR}$ since our
explicit numerical calculations indicate that, perhaps purely
coincidentally, $\rho_c \sim h/e^2$ ( again within a factor of $2-3$)
is also true for $\rho_c = \rho_c^{per}=\rho(n=n_c^{per})$. We discuss
the issue of localization versus percolation more below in the context
of comparing theory and experiment. 

There is one particular experimental finding, however, which can only
be explained by the quantum Ioffe-Regel theory with the classical
percolation theory failing completely. The dependence of $n_c$ on an
applied parallel magnetic field, which is widely reported
experimentally, can only be explained correctly by the Ioffe-Regel
theory and not at all by the percolation theory.

\subsection{Theory and experiment}

An important issue is how our theoretically calculated $n_c$ compares
with the observed experimental dependence of the critical density on  
various system parameters such as ``maximum" mobility, temperature,
external magnetic field, valley degeneracy, effective mass, etc. In
this respect (i.e., when compared with experimental findings), the
Ioffe-Regel criterion describing 2D MIT as a crossover phenomenon
seems to be in much better agreement (both qualitative and
quantitative) with observations than the percolation theory. In
particular, the fact that the 2D MIT behavior (specifically, the value
of $n_c$ itself) depends on an applied parallel magnetic field is
difficult to reconcile with the percolation transition which gives a
nominally density-independent explicit value of $n_c = n_c^{per}
\approx 0.1 \sqrt{n_i}/d$. The Ioffe-Regel theory by contrast
correctly predicts an increasing $n_c$ with the applied parallel field
(i.e., a field-induced suppression of the metallic phase much
discussed in the literature \cite{magnetic,hwangmagnetic} on 2D MIT)
arising from the 
spin-polarization of the 2D system. Similarly, the valley degeneracy
dependence of $n_c$ (and its equivalence to the spin degeneracy
dependence), which has been experimentally demonstrated
\cite{shayegannp}, is very naturally explained in the Ioffe-Regel
theory as arising from the variation in screening due to the
modification in the density of states, whereas it has no
explanation within the percolation theory. 
We see no obvious way of incorporating the applied magnetic field
effect in the percolation picture, and thus it appears that 2D MIT, at
least for the 2D systems manifesting a strong magnetic field
dependence of $n_c$, is incompatible with the percolation transition.

The most important experimental parameter determining the crossover
density $n_c$ is, of course, the sample quality (or disorder) as
discussed throughout this article. We emphasize that although $n_c$
obviously increases with increasing disorder, and this is the key
reason for the 2D MIT phenomena manifesting itself only in the 1990s
when sufficiently high-quality 2D systems could be studied with
sufficiently low values of $n_c$, there does not, in
principle, exist a simple relationship between $n_c$ and the sample
mobility $\mu$ (at a density $n \gg n_c$). The reason for this is that
the sample disorder is minimally determined by at least two
independent parameters ($n_i$ and $d$), and therefore it is, in
principle, allowed for $n_c$ and $\mu_m$ [with $\mu_m = \mu (n=n_m)$
  where $n_m \gg n_c$ is some specific high density] to be completely
independent parameters. Thus, in principle, a  
 sample with very high $\mu_m$ could have much higher $n_c$ than
 another sample with low $\mu_m$ although it is probably not very
 likely.

 With the above caveat in mind we can, however, obtain from the
 Ioffe-Regel (or percolation) theory how $n_c$ varies with $n_i$ and
 $d$ separately, and we can also calculate how the mobility $\mu(n)$
 varies with $n_i$ and $d$ as well as carrier density
 $n$. \cite{dassarmaprb2013} Therefore, the disorder dependence of $n_c$
 is completely specified in our theory through the two disorder
 parameters $n_i$ and $d$. Assuming a fixed $d$, we can convert the
 $n_i$-dependence of $n_c$ to an effective dependence on the mobility 
at some high carrier density, finding, $n_c \sim \mu_m^{-0.7}$ in the
Ioffe-Regel theory and $n_c \sim \mu_m^{-0.5}$ in the percolation
theory. The fact that Sarachik already pointed out more than 10 years
ago \cite{sarachikepl} an empirical relationship, $n_c \sim
\mu_m^{-0.7}$, which is in agreement with the Ioffe-Regel theory, is a
strong agreement in favor of the Ioffe-Regel theory.

Assuming $k_F d \ll 1$, we obtain theoretically $I_c^{IR} \sim n_id^0$
and $n_c^{per} \sim n_i d^{-1/2}$ whereas for $k_Fd \gg1$, $n_c^{IR}
\sim (n_i/d)^{2/3}$ and $n_c^{per} \sim (n_i/d)^{1/2}$. In principle,
these asymptotic dependence on $n_i$ and $d$ can be explicitly checked
experimentally, but we know of no detailed experimental study of the
critical density on the microscopic parameters defining the disorder.  

At this stage, the most convincing agreement between our theory for
$n_c$ and experiment comes from (1) $n_c^{IR} \sim \mu_m^{-0.67}$ type
behavior noted earlier empirically \cite{sarachikepl}; (2) the
parallel field induced enhancement of $n_c^{IR}$;
\cite{magnetic,hwangmagnetic} (3) the dependence of $n_c^{IR}$ on the
valley degeneracy and its equivalence to the spin-degeneracy
dependence \cite{shayegannp}. We note that all three properties
mentioned here favor the 2D MIT being a strong localization induced
crossover phenomenon as determined by the Ioffe-Regel criterion (in
contrast to the classical percolation transition). In this
context, we must mention one small (but significant) remaining
discrepancy between the Ioffe-Regel theory and the experimental
finding on 2D MIT. The Ioffe-Regel theory predicts that the  critical
resistivity $\rho_c = \rho(n_c)$ at the transition must necessarily
obey the inequality 
\begin{equation}
\rho_c^{IR} \leq \frac{h}{e^2} \frac{2}{g_sg_v},
\end{equation}
which means that even if the spin and valley degeneracy are lifted
$\rho_c < h/e^2 \approx 25,6000\Omega$.  
Experimentally, this inequality is obeyed almost universally with the
most important exception being the original Si-MOSFET data of
Kravchenko {\it et al.} who consistently found $\rho_c^{Si} \approx
1.5 h/e^2$. We have no way of explaining $\rho_c > h/e^2$ (at least at
$T=0$) within the Ioffe-Regel theory. One possibility is that $\rho_c
> h/e^2$ is a finite temperature effect, and $\rho_c(T \rightarrow 0)$
approaches $h/e^2$, but we simply do not know if this is true or not. 
We emphasize, however, that the vast majority of 2D MIT data are
consistent with $\rho_c$ values obtained from the Ioffe-Regel theory,
and thus the critical resistivity issue may not be a particularly
important problem for the Ioffe-Regel theory, particularly since the
extrapolated value of $\rho_c(T\rightarrow 0)$ is not easy to
ascertain experimentally from finite temperature transport
measurements.

\subsection{Transition versus crossover}

We have studied the 2D MIT as a crossover in this work (either
described by Ioffe-Regel criterion or by percolation), not as a true
localization quantum phase transition since two is established to be
the lower critical dimensions for the Anderson localization phenomenon
\cite{belitzrmp1994}, both for noninteracting electrons in a
disordered system \cite{abrahams1979,wegnerzp1979} and in the presence
of disorder and interaction \cite{finkel}. We consider the metallic
phase (for $n>n_c$) to be an effective metal which at $T=0$ will be
insulating in an infinite system. The metal-to-insulator crossover in
our theory arises from the strong modification in the effective
screened Coulomb disorder which becomes very strong as the carrier
density is lowered, leading to the $k_F l = 1$ condition defining the
MIT crossover point. The fact that our calculated $n_c$ is in
qualitative agreement with experimental observations is persuasive
evidence in support of 2D MIT being a crossover phenomenon, but our
theory can shed no light on the theoretical question of whether quantum
criticality in playing a role in this problem or not.  
In particular, we emphasize that we have no way of ruling out the 2D
MIT as a true quantum phase transition since this issue is simply
beyond the scope of our work, which treats the problem manifestly as a
crossover phenomenon described by the Ioffe-Regel criterion.

\section{conclusion}

Assuming 2D MIT to be a crossover phenomenon from a weakly localized
effective metallic phase to a strongly localized
insulating state, we have developed a theory for the critical density
for the transition from the higher-density effective metallic phase to
the lower-density strongly localized insulating phase. The calculated
critical density based on the well-known Ioffe-Regel criterion for
strong localization is in qualitative agreement with experimental
observations on 2D MIT with respect to its dependence on disorder,
applied parallel magnetic field, valley degeneracy, and materials
parameters. 

Our main findings are the following: (1) The critical density $n_c$
for the 2D MIT crossover varies with the maximum sample mobility
$\mu_m$ (measured at some high carrier density $n \gg n_c$) according
to the approximate scaling law, $n_c \sim \mu_m^{-\gamma}$, with the
exponent $\gamma \approx 0.7$ (for screened Coulomb disorder) and 1
(for purely zero-range $\delta$-function disorder) as derived from the
Ioffe-Regel criterion and $\gamma =0.5$ (for all disorder) as derived
from the semiclassical percolation theory. 
(2) The Ioffe-Regel criterion predicts an
enhancement of $n_c$ with decreasing spin-degeneracy, as, for example,
in the presence of an applied parallel magnetic field inducing
spin-polarization in the system provided that the critical density
$n_c$ is fairly low at zero spin-polarization (so that the condition
$q_{TF} \agt 2k_F$ or $q_{TF} \gg 2k_F$ is satisfied). By contrast,
the percolation theory predicts no dependence of $n_c$ on the spin
degeneracy. (3) The Ioffe-Regel criterion predicts an enhancement of
$n_c$ with decreasing valley degeneracy in the system as, for example,
could be induced by applying a suitable external strain. In fact, the
Ioffe-Regel criterion predicts that the dependence of the critical
density $n_c(g_s,g_v)$ on the spin- and valley-degeneracy to be
approximately equivalent, i.e., increasing (decreasing) $g_s$ or
$g_v$ decreases (increases) $n_c$ if all other parameters are
fixed. This mutually equivalent spin and valley-degeneracy dependence
of $n_c$ arises from 
the 2D screening being dependent on $g_s$ and $g_v$ equivalently since
the density 
of states is proportional to $g_s g_v$. One direct prediction of the
Ioffe-Regel theory is thus that the most metallic (insulating) situation
will manifest itself for the largest (smallest) values of the product
$g_s g_v$, and therefore, $n_c$ will be the smallest (largest) for the
largest (smallest) values of $g_s g_v$ in the system. The percolation
theory predicts the 2D MIT phenomena to be completely independent of
$g_s$ and/or $g_v$, and thus does not in any way predict any
dependence of $n_c$ on spin- or valley-polarization.
We mention that all three of these theoretical findings based on the
Ioffe-Regel criterion are in good qualitative and semiquantitative
agreement with experimental results on 2D MIT whereas the predictions
of the percolation theory -- in particular, the $n_c \propto
\sqrt{\mu_m}$ dependence and the lack of dependence of $n_c$ on $g_s$
and $g_v$ -- are in disagreement with the empirical evidence.

In addition to the above qualitative physical results following
directly from our theory, we also find that the actual quantitative
values of $n_c$ calculated on the basis of the Ioffe-Regel criterion
to be consistent with the experimental results. In particular, we find
that the Ioffe-Regel criterion gives the following approximate
critical density values for typical high-mobility 2D systems studied
in the existing 2D MIT literature \cite{abrahamsrmp,kravchenko,ssc,spivakrmp,dassarmarmp}: $n_c \approx 10^{11}$
cm$^{-2}$ (for Si MOSFETs); $10^{10}$ cm$^{-2}$ (for p-GaAs); $10^9$
cm$^{-2}$ (for n-GaAs). By contrast, we find that the lower-quality Si
MOSFETs studied extensively during the 1970s and early 1980s
\cite{andormp} should typically have $n_c \approx 10^{12}$ cm$^{-2}$
according to the Ioffe-Regel criterion. This high value of $n_c$, with
a corresponding Fermi temperature $\sim 73$ K, is not only consistent
with the experimental $n_c$ values found in older MOSFETs
\cite{andormp} with lower values of maximum mobility ($\mu_m < 5000$
cm$^2$/Vs), but also provides an obvious explanation for why the 2D
MIT phenomenon could only be observed after very high mobility ($\mu_m
\agt 20,000$ cm$^2$/Vs) Si MOS samples become available in the 1990s.
\cite{kravchenko1995} A high value of $n_c$ with a concomitant high
value of Fermi temperature makes it impossible \cite{sarmaprb2004}
for the 2D 
effective metallic phase (for $n>n_c$) to manifest any temperature
dependence in its resistivity arising purely from an electronic
mechanism.

Several open questions remain for future investigations. Although our
main conclusion is that the 2D MIT (at least at very low temperatures)
is strong-localization induced crossover phenomenon as constrained by
an Ioffe-Regel type quantum interference condition, the role of
percolation in the inhomogeneous potential fluctuation
landscape in affecting the crossover behavior remains unclear. 
For Si-MOSFETs, where the random charged impurities are located at the
Si-SiO$_2$ interface close to the 2D electron system, it is hard to
see how and why percolation could be relevant, but in modulation-doped
GaAs structures, where the dopants are far away from the electrons,
percolation could conceivably be relevant. One possibility we have
speculated about is that the transition itself crosses over from being
percolation like at higher temperatures to being Ioffe-Regel-like at
lower temperatures as quantum tunneling and quantum interference
become effective. But these are all mere speculations, and we do not
have a theory combining percolation and strong localization crossover,
which remains an important open issue in the long-range 
fluctuating potential landscape of Coulomb disorder. In the current
work, we have only compared localization and percolation crossovers as
distinct physical processes,
concluding that the experimental observations are more consistent with
the Ioffe-Regel localization crossover.

Another open question (and an important shortcoming of our theory) is
that the theory developed in this paper approaches the transition
(i.e., the crossover) from the higher-density effective metallic side
(with decreasing density to approach the transition) using Boltzmann
transport theory to treat the screened disorder induced
carrier scattering. An equivalent theory from the lower-density
insulating side (with increasing density to approach the transition)
is highly desirable, but is out of scope for our work, and in fact,
there is no good idea in the literature about how to approach the
transition from the insulating side where the whole concept of a
quantum mean free path becomes inapplicable (and therefore the
Ioffe-Regel criterion is useless). Such a theory from the insulating
side, if available, could be a compelling consistency check for the
calculated critical density if the same crossover point is reached
theoretically from either direction.

Another issue with our theory, in spite of its good qualitative
agreement with essentially all aspects of 2D MIT phenomenology, is
that our calculated critical resistivity (i.e., the 2D resistance at
the crossover critical density) is only in approximate quantitative
agreement with experiments. This may not be a serious problem since
the crossover nature of the transition makes it problematic to define
a unique zero-temperature critical resistivity (particularly since the
2D resistivity is strongly temperature dependent around the critical
density), and it is likely that a proper extrapolation of the
experimental data to zero temperature would be in reasonable agreement
with our theory since the disagreement is mainly in Si-MOSFETs and is
by less than a factor of two. 
More experimental work is necessary to settle the question for the
precise value of the $T=0$ critical resistivity in 2D MIT in various
systems.

Our theory provides a good qualitative explanation for the dependence
of the critical density on disorder, applied parallel
magnetic field, spin and valley degeneracy, and materials parameters
indicating that the Ioffe-Regel criterion, in all likelihood, captures
the essential features of the transition between the high-density
effective (weakly localized) metallic phase and the low-density
strongly localized insulating phase in 2D semiconductors. 
The issue of whether 2D MIT is or is not a true $T=0$ quantum phase
transition as well as whether or how electron-electron interaction
\cite{finkel} 
beyond screening affects the transition, however, still remaining open
as a theoretical question 
for future work.
What we have established in this work through extensive calculations
is that the application of the empirical Ioffe-Regel criterion, which
is often used in the literature for a semi-quantitative description of
the Anderson localization transition in three-dimensional systems, to
the phenomenon of the apparent two-dimensional metal-insulator
transition provides a critical density which agrees well with existing
experiments in describing the characteristic dependence of the
critical density on disorder, mobility, temperature, and magnetic
field, indicating that the observed 2D MIT phenomenon is likely to be
a crossover between a weakly localized 2D metal and a strongly
localized Anderson insulator.

\section*{Acknowledgements}

This work is supported by LPS-CMTC and Basic Science Research Program
through the National Research Foundation of Korea Grant funded by the
Ministry of Science, ICT \& Future Planning (2009-0083540).

\end{document}